\documentclass[aps,prb,twocolumn,a4paper,showpacs,superscriptaddress,groupedaddress]{revtex4-1} 
\usepackage{amsmath}
\usepackage{amssymb}
\usepackage{bm}
\usepackage[usenames]{color}
\usepackage[normalem]{ulem}
\usepackage[dvips]{graphicx} 

\definecolor{myorange}{rgb}{0.7,0.5,0.0}
\definecolor{mygreen}{rgb}{0.0,0.7,0.0}
\definecolor{purple}{rgb}{0.75,0.0,1.0}
\definecolor{mahogany}{rgb}{0.8,0.0,0.0}

\newcommand{\tr}[1]{\textcolor{black}{#1}}

\newcommand{\tb}[1]{\textcolor{black}{#1}}
\newcommand{\tm}[1]{\textcolor{black}{#1}}

\newcommand{\tbb}[1]{\textcolor{black}{#1}}

\newcommand{\tc}[1]{\textcolor{black}{#1}}

\newcommand{\txr}[1]{\textcolor{black}{#1}}

\newcommand{\txm}[1]{\textcolor{black}{#1}}

\newcommand{\sP}{\mathcal{P}}
\newcommand{\sL}{\mathcal{L}}

\newcommand{\la}{\langle}
\newcommand{\ra}{\rangle}

\newcommand{\Ns}{N_{\text{s}}}

\begin{document}
  \title{{Effective Hamiltonian for cuprate superconductors derived from\\
multi-scale \textit{ab initio} scheme with level renormalization}}
  \author{Motoaki Hirayama$^{1)}$, Takahiro Misawa$^{2)}$, Takahiro Ohgoe$^{3)}$, Youhei Yamaji$^{3)}$, and Masatoshi Imada$^{3)}$}
  \affiliation{$^{1)}$RIKEN Center for Emergent Matter Science, Wako, Saitama 351-0198, Japan}
  \affiliation{$^{2)}$Institute for Solid State Physics, University of Tokyo, Kashiwanoha, Kashiwa, Chiba, Japan}
  \affiliation{$^{3)}$Department of Applied Physics, University of Tokyo, 7-3-1 Hongo, Bunkyo-ku, Tokyo 113-8656, Japan}
  
\begin{abstract}
Three-types (three-band, two-band and one-band) of effective Hamiltonians for the HgBa$_2$CuO$_4$ and three-band effective Hamiltonian for La$_2$CuO$_4$ are derived beyond the level of the constrained-GW approximation combined with the self-interaction correction (cGW-SIC) derived in Hirayama {\it et al.} Phys. Rev. B \textbf{98}, 134501 (2018) by improving the treatment of the interband Hartree energy.
The charge gap and antiferromagnetic ordered moment show good agreement with the experimental results when the present effective Hamiltonian is solved, indicating the importance of the present refinement.
The obtained Hamiltonians will serve to clarify the electronic structures of these copper oxide superconductors and to elucidate the superconducting mechanism.
\end{abstract}

\maketitle

\section{Introduction}
Mechanism of high temperature superconductivity in copper oxide superconductors discovered 
more than thirty years ago\cite{bednortz} is still under active debates. 
One of the reasons of the controversies is severe competitions of completely different orders, 
particularly, $d$-wave superconductivity, antiferromagnetism and charge inhomogeneous states 
such as charge/spin stripe-ordered states suggested by experiments~\cite{tranquada,yamada,ghiringhelli,tabis,comin,forgan,peng,eduardo,fujita,mesaros,keimer,comin16} as well as 
by highly accurate numerical studies on theoretical models 
such as the Hubbard model\cite{white00,capone06,misawa14,corboz14,otsuki14,zhao17,zheng17,ido18},
while they are still controversial. Therefore, more quantitative {\it ab initio} studies are 
needed based on the realistic parametrization of the cuprate superconductors 
to reach conclusive, and quantitative understanding of the mechanism.

Recently, several first principles effective Hamiltonians 
for low-energy degrees of freedom of electrons near the Fermi level 
in the cases of Hg based and La based cuprate superconductors have been 
derived after eliminating the high-energy electronic 
degrees of freedom far from the Fermi level\cite{hirayama18}, 
based on the multi-scale {\it ab initio} scheme for correlated electrons 
(MACE)~\cite{imadamiyake10,hirayama13,hirayama17},
which is expected to be the basis of quantitative realistic studies of the cuprate 
superconductors without adjustable parameters. The derivation of the effective 
Hamiltonians is based on the constrained GW (cGW) calculation, 
where the exchange correlation energy and the Hartree energy 
in the density functional theory (DFT) in the level of 
the local density approximation (LDA) is carefully removed to 
exclude the double counting of the Coulomb interaction in 
the low-energy effective Hamiltonians. 
Other attempts to determine parameters of effective Hamiltonians 
were also reported~\cite{hybertsen,anisimov,sakakibara}.

In this paper, we propose a more accurate and realistic description of the 
low-energy effective Hamiltonians by taking account effects called energy 
{\it level renormalization (LR)} of the orbitals consisting of the 
low-energy effective Hamiltonians. In the present framework, 
effects of the Hartree energy between the low-energy 
orbitals contained in the effective Hamiltonians and 
the high-energy orbitals outside of them already eliminated in the effective 
Hamiltonians are calculated more accurately. This Hartree energy contribution 
has been of course taken into account in the GW level. However, when we solve the 
low-energy effective Hamiltonians more accurately beyond the GW, the charge density 
is improved and the Hartree energy is modified. This correction is not taken 
into account in Ref.\onlinecite{hirayama18} and generate the LR.

We further take into account the feedback from the LR to 
the GW global band structure. By using the renormalized global 
band structure, we derive an improved effective Hamiltonian 
using the cGW
calculation. By this correction, 
we show that the level distance in the low-energy orbitals 
is renormalized to smaller values and the resultant 
enhanced mutual screening between these orbitals 
drives the effective interaction weaker in the low-energy Hamiltonians.
We show that the improved Hamiltonian well reproduces the charge gap and antiferromagnetic ordered moment of the experimental results in the mother materials.

In Sec. II we show the method of the improved downfolding.
The three effective Hamiltonians for HgBa$_2$CuO$_4$ are derived in Sec. III.A.
The result obtained by the variational Monte Carlo method (VMC)~\cite{Gros,TaharaVMC} to incorporate the feedback is also shown in Sec. III.A.
Three-band effective Hamiltonians for La$_2$CuO$_4$ are derived in Sec. III.B.
We summarize the paper in Sec. IV.

\section{Method}
\subsection{Downfolding method}
\subsubsection{cGW-SIC}

The aim of MACE is to {derive} an \textit{ab initio} effective Hamiltonian
for the low-energy degrees of freedom from the whole 
band structure of all {electronic} degrees of freedom, 
particularly for strongly correlated electron systems.
The effective Hamiltonian in the low-energy space is 
given in the form of extended Hubbard-type Hamiltonian
without any adjustable parameters as,
\begin{eqnarray}
\mathcal{H}_{\text{eff}} ^{\text{cGW-SIC}}= \sum_{ij} \sum_{\ell_1 \ell_2\sigma }&&
t^{\text{cGW-SIC}}_{\ell_1 \ell_2\sigma}(\bm{R}_i-\bm{R}_j) d_{i\ell_1\sigma} ^{\dagger} d_{j \ell_2\sigma} \nonumber \\
+ \frac{1}{2} \sum_{i_1i_2i_3i_4} \sum_{\txr{\ell_1 \ell_2 \ell_3 \ell_4} \sigma \eta \rho \tau}  
&\biggl\{& W_{ \ell_1 \ell_2 \ell_3 \ell_4\sigma \eta \rho \tau }^r(\bm{R}_{i_1},\bm{R}_{i_2},\bm{R}_{i_3},\bm{R}_{i_4}) \nonumber \\
&&d_{i_1 \ell_1\sigma}^{\dagger}d_{i_2 \ell_2\eta} d_{i_3 \ell_3\rho}^{\dagger} d_{i_4 \ell_4\tau}\biggl\},
\label{Hamiltonian0}
\end{eqnarray}
where $d_{i\ell\sigma} ^{\dagger}$ ($d_{i \ell\sigma}$) is the creation
(annihilation) operator of an electron for the $\ell$th Maximally localized Wannier function (MLWF) with spin $\sigma$ centered at unit cell $\bm{R}_i$.
The terminology ``extended Hubbard Hamiltonian" is used in this paper as the lattice fermion Hamiltonian containing longer-ranged transfers as well as longer-ranged and/or off-diagonal Coulomb interactions to represent first principles parameters accurately beyond the simple Hubbard model containing only the onsite interaction and the nearest-neighbor transfer.
Here, the single particle term is represented by 
\begin{equation}
t^{\text{cGW-SIC}}_{ \ell_1 \ell_2\sigma}(\bm{R})= \langle \phi _{ \ell_1\bm{0}}|{H}^{\text{cGW-SIC}}_{K}|\phi _{ \ell_2\bm{R}} \rangle, 
\label{cGW-SICK}
\end{equation}
and  the interaction term is given by 
\begin{eqnarray}
W_{ \ell_1 \ell_2 \ell_3 \ell_4\sigma \eta \rho \tau }^r(\bm{R}_{i_1},\bm{R}_{i_2},\bm{R}_{i_3},\bm{R}_{i_4}) \nonumber \\
= \langle \phi _{ \ell_1\bm{R}_{i_1}}\phi _{ \ell_2\bm{R}_{i_2}}|{H}^{\text{cGW-SIC}}_{W^r}|\phi _{ \ell_3\bm{R}_{i_3}}\phi _{ \ell_4\bm{R}_{i_4}} \rangle ,
\label{cGW-SICW}
\end{eqnarray}
where $\phi _{ \ell\bm{R}_{i}}$ is the $\ell$th MLWF centered at $\bm{R}_i$.
In the previous approach\cite{hirayama13,hirayama17,hirayama18}, 
the one-body term ${H}^{\text{cGW-SIC}}_{K}$ and the 
2-body term ${H}^{\text{cGW-SIC}}_{W^r}$ were calculated by the 
cGW with the self-interaction correction (SIC) and 
the constrained random phase approximation (cRPA), 
respectively, using the Green's function of the band structure of all degrees of freedom.
It should be noted that all the parameters in 
Eq.(\ref{Hamiltonian0}), namely $t^{\text{cGW-SIC}}$ and $W^r$, are given 
from the first principles calculation without any adjustable parameters. 
In this research, we follow the basic strategy of MACE 
and use the whole band structure {obtained by} the GW approximation (GWA) beyond the DFT
to derive the {\it ab initio} Hamiltonian.
As is widely known, the band gap, or more generally, energy difference between low-energy bands, is underestimated in the DFT scheme using the LDA,
while it is improved by the GW method~\cite{louie86}.
Both the one-body and two-body parts in the {\it ab initio} Hamiltonian, therefore, are also
expected to be more accurate by using the GW method.

In the cGW\cite{hirayama13,hirayama17}, the band dispersion is determined 
from the self-energy and the polarization by excluding the 
contribution within the low-energy degrees of freedom to 
remove the double counting. These contributions from the low-energy degrees of freedom
are taken into account afterwards when the low-energy effective Hamiltonian is solved
in the same way as the LDA+cRPA scheme based on the LDA Kohn-Shame Hamiltonian. 
However, in contrast to the LDA+cRPA, the cGW method can explicitly exclude the double counting of
the exchange correlation energy in the effective Hamiltonian
because the contributions from high- and low-energy degrees of freedom to the exchange correlation energy
can be disentangled in the GW scheme~\cite{aryasetiawan09} while their contributions cannot be separated in the DFT.
Furthermore, in the GW-based scheme, the electron correlation from the degrees of freedom outside of 
the effective Hamiltonian is better taken into account than the LDA~\cite{hirayama18}. 
The self-interaction included in the LDA is also removed by 
the self-interaction correction (SIC) that subtracts the 
Hartree energy estimated from the LDA charge density
of the Wannier orbitals in the low-energy effective Hamiltonian.
The double counting of Hartree energy is 
subtracted when the effective Hamiltonian is solved.  
Furthermore, the frequency dependent part of the interaction 
ignored in the low-energy Hamiltonian 
is taken into account as the renormalization factor in the one-body part.

\subsubsection{Error in cGW-SIC}

Even with this cGW-SIC formalism, an important correction to 
the Hartree energy contribution is missing. When the low-energy effective Hamiltonian 
is solved, the high-energy degrees of freedom are already traced out, 
and the ground state is determined only from the energy of the low-energy degrees of freedom.
In the solution, the spatial distribution of the electron density 
(the primary part is the electron occupation in the Wannier orbitals 
in the low-energy degrees of freedom) changes in general from that in the GW (or DFT).
This change in the electron density makes a difference in the 
Hartree interaction between the low- and high-energy degrees of freedom,
which is not taken into account in the low-energy solver.
However, this difference of the interband Hartree energy can be substantial, 
because the number of high-energy bands are large and, thus, a 
small change in charge density may induce a large change in the interband Hartree energy.

\subsubsection{Rigidity of orbital occupation}

The number of degrees of freedom and the scale of total energy are greatly different between the all-electron calculation and the low-energy effective Hamiltonian.
The electron density in the all-electron calculation is determined by the bare Coulomb interaction of about 20 eV at the on-site and several eV at most at off-sites.
On the other hand, the electron density in the low-energy effective Hamiltonian is determined only by the screened interaction between the low-energy degrees of freedom, which is one order of magnitude smaller than the bare Coulomb interaction.
In the low-energy effective Hamiltonian, the high-energy degrees of freedom is traced out, and it is impossible to account for the change in the total energy of the high-energy degrees of freedom due to the change of the electron density of the low-energy degree of freedom.
Since the change in the charge distribution causes significant increase in the Hartree energy, the charge distribution is hardly affected by further improving accuracy of the {\it ab initio} methods (see Appendix \ref{appendix_rigidity}).
The interband energy in the Hartree level determined from the 
global electronic structure is actually properly calculated 
in the GW energy and the resultant stable charge distribution is reliable.
In fact, the Hartree level of energy and resultant charge density 
is estimated both by the LDA and GW with very similar values.
For example, the occupation numbers for the Cu $x^2-y^2$ and the O $2p$ orbitals in the LDA/GW are 1.450 and 1.775/1.437 and 1.781 in the Hg system, 1.396 and 1.802/1.350 and 1.825 in the La system, respectively, and the LDA and GW show no appreciable difference.
The charge density may not be affected even when more accurate {\it ab initio} treatments are used.
This means that the orbital occupation is rigid and should be 
fixed at the values of the GW (or similar LDA) results 
in the solution of the low-energy solver. 
This rigidity of the orbital occupation is expected to be more accurate 
if the Wannier orbitals belong to different atoms, 
because the Hartree energy is expected to be very different 
for orbitals belonging to different atoms and even a 
small redistribution of the charge in the low-energy 
orbitals results in large cost of interband Hartree energy.

\subsubsection{Chemical potential shift}
Then a better solution of the low-energy solver is obtained by shifting the chemical potential of each orbital in the effective low-energy Hamiltonian to adjust and reproduce the occupation in each orbital to the value given by the GW. 
We call the method to use the effective Hamiltonian simply obtained by such a shift of the chemical potential to the cGW-SIC Hamiltonian, cGW-SIC$+\Delta\mu$.

We note here about a subtlety of the cGW-SIC$+\Delta\mu$.
First, the ground state of the effective Hamiltonian obtained by the low-energy solver
may show spontaneous symmetry breaking while the GW solution is paramagnetic: The ground state of
the effective Hamiltonian at half filling obtained by the VMC has antiferromagnetic order,
while the paramagnetic ground state is obtained by the present GW calculation.
This difference in the ground state character may introduce the possible correction arising from the exchange splitting effect, which is taken into account in the VMC result while it is not in the GW energy.
Another subtlety is the off-diagonal part of the density fitting.
Although it is a secondary effect, the Hartree energy contains not only the diagonal part of density
($d_i^{\dagger}d_i$ and $d_j^{\dagger}d_j$ ) but also the off-diagonal part ($d_i^{\dagger}d_j$ and $d_j^{\dagger}d_i$)
in the atomic orbital basis of $d_i$ and $d_j$.
This off-diagonal part may also be adjusted between the GW and the VMC results.
These secondary effects will be discussed in a future publication.

\subsubsection{Renormalized level feedback}

\begin{figure}[h]
\centering 
\includegraphics[clip,width=0.4\textwidth ]{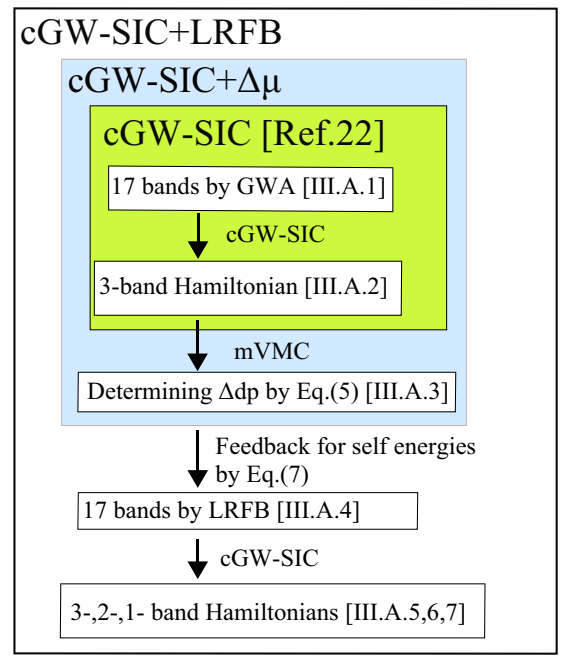} 
\caption{(Color online) 
Overview of cGW-SIC+LRFB method. 
As an example, we show the calculation flow
of cGW-SIC+LRFB method for 
the cuprate superconductors.
}
\label{outline}
\end{figure}

The renormalized level determined in the cGW-SIC$+\Delta\mu$ can further be used to improve the full GW electronic structure.
We call this improvement level renormalization feedback (LRFB).
In the LRFB, the chemical potential shifts are added to the self-energy to update the full GW Green's function,
where we call the updated one
GW+LRFB Green's function. 
For better self-consistency, we use the GW+LRFB Green's function to perform the cGW-SIC calculations again.
In this paper, we employ the level-renormalized revised effective 
Hamiltonians by taking into account the feedback effect in this way. We name this scheme cGW-SIC+LRFB.
The outline of cGW-SIC+LRFB method is illustrated in Fig.~\ref{outline}.
The present procedure can be self-consistent by repeating
the LRFB in
the cGW-SIC$+\Delta\mu$
until the the cGW-SIC+LRFB 
effective Hamiltonian coverges (as shown in Fig.~\ref{outline}),
while it is beyond the scope of the present paper.
We describe details of cGW-SIC+LRFB below.

\begin{figure}[h]
\centering 
\includegraphics[clip,width=0.3\textwidth ]{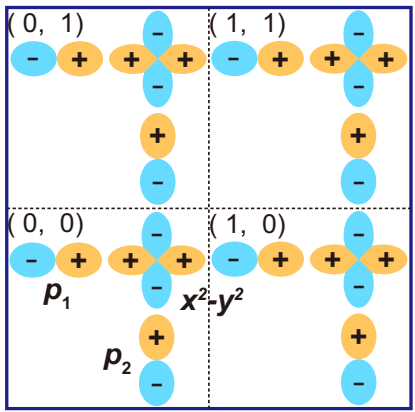} 
\caption{\tm{(Color online) 
Position of the Cu $x^2-y^2$ and O $2p$ orbitals 
and definition of the $2\times 2$ sublattice in the VMC for the three-band Hamiltonian.
{Phases of the orbitals are denoted by $\pm$.}
}}
\label{dpp2x2}
\end{figure}

The cGW-SIC$+\Delta\mu$ and cGW-SIC+LRFB are expected to reach a 
non-negligible improvement of the low-energy effective Hamiltonian. 
To demonstrate the improvement,
here, we take an example of the three-band Hamiltonian 
for the cuprates derived in Ref.\onlinecite{hirayama18}, where the 
Wannier orbitals of $d$ and $p$ belong to different atoms, 
namely Cu and O, respectively. If the LRFB is applied, the level difference 
between the oxygen $p_{\sigma}$ and the copper $d_{x^2-y^2}$ orbital 
decreases in comparison to the GW results, as we show later.
This also means that the effective one-band Hamiltonian for the anti-bonding band 
resulting from the hybridized $d_{x^2-y^2}$ and $p_{\sigma}$ orbitals 
has to be further improved because of the level shift and 
increased screening from the $p_{\sigma}$ orbital.
More precisely, after 
the hybridization of the Cu$d$ and O$p$ orbitals, the GW results are given by the bonding, 
non-bonding, and anti-bonding bands. Since the LR leads to 
stronger hybridization and screening effects arising from the bonding and non-bonding bands, 
they may make the effective interaction of the one-band Hamiltonian weaker.
For example, as we will show later, nearest-neighbor hopping $t(1,0,0)$, the on-site screened Coulomb interaction $U$, and $U/t(1,0,0)$ for the one-band Hamiltonian of HgBa$_2$CuO$_4$ are $-0.461$, $4.37$, and $9.5$ in the cGW, and $-0.509$, $3.85$ and $7.6$ in the GW+LRFB.

\subsection{Multi-scale correction for occupation number of low-energy effective {Hamiltonian}}

The improved transfer integral with the correction of the chemical potential $\tilde{t}$ is 
\begin{equation}
\tilde{t}^{\text{cGW-SIC}}_{ \ell_1 \ell_2\sigma}(\bm{R})
=t^{\text{cGW-SIC}}_{ \ell_1 \ell_2\sigma}(\bm{R})+\Delta\mu _{ \ell_1 \sigma}(\bm{0})\delta _{ \ell_1 \ell_2}, 
\label{cGW-SICK+on}
\end{equation}
where $\Delta\mu$ is the chemical potential shift to reproduce the occupation number of each Wannier orbital by the GW calculation, even after solving the effective Hamiltonian by an accurate low-energy solver.

Instead of DFT, in the GWA, the Green's function and various physical quantities such as the self-energy are calculated in a self-consistent manner based on the Hedin's equation.
In actual calculations, Hartree interaction is often not updated during the self-consistent calculation, but it is empirically known that the band structure is improved compared to that in the DFT.
Therefore, in this study, we employ the electron density of each Wannier orbital in the GWA in Eq. (\ref{cGW-SICK+on}) as the density to be reproduced in the solution of the low-energy effective Hamiltonian.

\subsection{Calculation by low-energy solver to correct orbital occupation number}

We have several possibilities for the choice of the low-energy solver when we solve the effective Hamiltonians derived by the cGW-SIC+LRFB.
Here, we employ the VMC method using the variational wave functions with various correlation factors and projection operators operated to pair product wave functions~\cite{TaharaVMC} as the low-energy solver.
The VMC is a method to optimize the variational parameters in the wave function
to reach a good variational many-body ground-state.

We replace $t^{\rm cGW\mathchar`-SIC}$ in the effective Hamiltonian (\ref{Hamiltonian0}) by $\tilde{t}^{\rm cGW\mathchar`-SIC}$ defined in Eq.(\ref{cGW-SICK+on})
and solve the modified effective Hamiltonian by sweeping the chemical potential $\Delta \mu$ of each orbital~\cite{hirayama17}.
Then we adjust the chemical potential of each orbital so that the VMC solution of the effective Hamiltonian, $|\psi\ra$, reproduces each orbital occupation obtained by the GWA, i.e., 
\begin{align}
\tr{n^{\rm VMC}_{\nu}=n^{\rm GW}_{\nu}},
\label{nadjust}
\end{align}
where $n_{\nu}^{\rm VMC}=N_{\rm s}^{-1}\sum_{i=1}^{N_{\rm s}}\sum_{\sigma=\uparrow,\downarrow}\la\psi|d_{i\nu\sigma}^{\dagger}d_{i\nu\sigma}^{\ }|\psi\ra$.

By employing the adjusted chemical potential suggested in the VMC result to satisfy Eq.(\ref{nadjust}), and shifting the chemical potential in the cGW-SIC Hamiltonian,
the cGW-SIC\tb{+}$\Delta\mu$ Hamiltonian is obtained.

\subsection{Downfolding with cGW-SIC+LRFB}
In the cGW-SIC+LRFB method, a static 1-body term ${H}^{\text{cGW-SIC}}_{K}$
is obtained from the dynamical 1-body term 
by renormalizing the frequency dependence using renormalization factors $Z_{\rm H}^{\rm cGW}$.
By multiplying the chemical  potential shift $\Delta\mu$ by $(Z_{\rm H}^{\rm cGW})^{-1}$,
the correction from the frequency dependence of
the dynamical 1-body term is taken into account.
The revised GW self-energy $\Sigma^{\text{LRFB}}(\omega )$ is then given by
\begin{equation}
\Sigma^{\text{LRFB}}(\omega ) = \Sigma ^{\rm GW} (\omega) + [Z_{\rm H}^{\rm cGW}(0)]^{-1} \Delta\mu .
\label{GW+LRFB}
\end{equation}
The second term in Eq.~(\ref{GW+LRFB}) is a contribution of correlation effect beyond the GWA.
The Hamiltonian in the GW+LRFB is
\begin{equation}
H = Z^{\text{LRFB}}(0)[ H^{\text{LDA}}-V^{\text{xc}}+ \Sigma^{\text{LRFB}}(0)],
\label{H:GW+LRFB}
\end{equation}
where $H^{\text{LDA}}$ is the Kohn-Sham Hamiltonian, $V^{\text{xc}}$ is the exchange correlation energy in the LDA results, and $ Z^{\text{LRFB}}$ is the renormalization factor of $\Sigma^{\text{LRFB}}$ at $\omega = 0$
\begin{equation}
Z^{\text{LRFB}}(0)= \biggl\{ I-\frac{\partial \text{Re}\Sigma^{\text{LRFB}}}{\partial \omega }\Big|_{\omega =0} \biggr\}^{-1}.
\label{Z:GW+LRFB}
\end{equation}
$Z^{\text{LRFB}}(0)$ is nearly the same as $Z^{\rm GW}(0)$ because the $\omega $ dependence of $\Sigma^{\text{LRFB}}$ originates from $\Sigma^{\text{GW}}$. 
In the Hamiltonian in the GW+LRFB,
the LR in the full GW level is mostly given by $Z^{\rm GW}(0)[Z_{\rm H}^{\rm cGW}(0)]^{-1} \Delta\mu$.
One might think that the LR in the full GW level 
is smaller than $\Delta\mu$ because 
$Q\equiv Z ^{\rm GW}(0)/Z_{\rm H}^{\rm cGW}(0)<1$.
However this is reasonable because after the cGW calculation, the LR is $\Delta\mu$ and the contribution from the low energy degrees of freedom gives further renormalization given by $Q$.

We use the corrected self-energy (\ref{GW+LRFB}) for the Green's function. Then  we perform the cGW-SIC again, which generates  cGW-SIC+LRFB Hamiltonian.

\subsection{Application to the cuprates}
In this paper, we apply the method to derive three types of effective Hamiltonian for the the cuprate superconductors HgBa$_2$CuO$_4$ and La$_2$CuO$_4$: (1) Three-band effective Hamiltonian consisting of the Cu $d_{x^2-y^2}$ and two O $2p_{\sigma}$ Wannier orbitals, (2) two-band Hamiltonian consisting of the Cu $d_{x^2-y^2}$ and Cu $d_{3z^2-r^2}$ Wannier orbitals and (3) one-band Hamiltonian for the anti-bonding band of hybridized Cu $d_{x^2-y^2}$ and O $p_{\sigma}$ Wannier orbitals.

In the present application to the cuprates, we apply the orbital level shift to the three-band cGW-SIC Hamiltonian in the form of  Eq.~(\ref{Hamiltonian0}) so that the relative level of O $2p_{\sigma}$ orbital is adjusted relative to the level of Cu  $d_{x^2-y^2}$ orbital.
To analyze the three-band Hamiltonian with the level shift,
we use the mVMC method.
In the mVMC calculation, we only consider the 
density-density type interactions ($W^r_{\ell_1 \ell_1 \ell_2 \ell_2 \sigma \sigma \rho \rho}(\bm{R}_{i_1},\bm{R}_{i_1},\bm{R}_{i_2},\bm{R}_{i_2})$) and
ignore the exchange term because their effects are small.

The present scheme is summarized in the following (see Fig.~\ref{outline}).
Following the treatment employed in Ref.\onlinecite{hirayama18}, the  effective Hamiltonian for the 17 bands near the Fermi level is first derived. For this purpose, the global band structure is obtained by the DFT with LDA. Then the Green's functions for the bands other than the 17 bands are fixed in this LDA form all through the calculations.  The band structure of the 17 bands are first derived from the self-energy of the 17 bands calculated from the one-shot full GW calculation.  Next, by using the GW Green's function for the 17 bands, the cGW-SIC calculation is performed to derive the effective Hamiltonian for the 17 bands with the one-body term obtained from the cGW and the interaction term using the cRPA.  Then from this Hamiltonian, the three types of effective Hamiltonians are derived as we detail below.  Up to here the procedure is the same as that employed in Ref.\onlinecite{hirayama18}.

By adding additional chemical potentials $\Delta\mu_d$ and $\Delta\mu_p$ for the Cu $d$ and O $p$ orbitals, respectively as parameters, $\Delta\mu_{dp}\equiv  \Delta\mu_p-\Delta\mu_d$ dependence of the orbital fillings is calculated by the mVMC for the three-band Hamiltonian. In general the orbital fillings in the mVMC solution are not the same as the full GW result if $\Delta\mu_{dp}=0$. Then the relative chemical potential $\Delta\mu_{dp}$ is shifted to the value so that the orbital fillings in the mVMC solution become the same as the full GW result.  
By employing this level shift to the cGW-SIC Hamiltonian, cGW-SIC+$\Delta\mu$ Hamiltonian is obtained.  By taking into account the effect of nonzero $\Delta\mu_{dp}$, we recalculate the cGW-SIC to rederive the effective cGW-SIC-LRFB Hamiltonian with the LRFB correction.

\subsection{Computational Conditions}
\subsubsection{Conditions for DFT, and GW}
For the crystallographic parameters, we employ the experimental results reported by Ref.~\onlinecite{Putilin} for HgBa$_2$CuO$_4$
and those reported by Ref.~\onlinecite{Jorgensen} for La$_2$CuO$_4$. 
We take
the lattice constants of the tetragonal unit cell as
$a=3.8782/3.7817$\AA 
and $c=9.5073/13.2487$\AA 
for the Hg/La compound.
In the Hg compound, the height of Ba atom measured from CuO$_2$ plane is $0.2021c$ and the apex oxygen height is $0.2940c$.
In the La compound, the La and apex oxygen heights measured from the CuO$_2$ plane are $0.3607c$ and $0.1824c$, respectively.
Other atomic coordinates are determined from the crystal symmetry.
Here, the crystallographic parameters of HgBa$_2$CuO$_{4+\delta}$ is employed because the mother compound is not available. The mother compound La$_2$CuO$_4$ has an orthorhombic symmetry and has slightly different lattice parameters from those listed above. We employ the tetragonal symmetry for the effective Hamiltonian with the crystallographic parameters of La$_{1.85}$Ba$_{0.15}$CuO$_4$ at 10K. We neglect this difference.

Computational conditions are as follows.
The band structure calculation is based on the full-potential
linear muffin-tin orbital (LMTO) implementation~\cite{methfessel}.
The exchange correlation functional is obtained by the local density approximation of the Ceperley-Alder type~\cite{ceperley}
and spin-polarization is neglected.
The self-consistent LDA calculation is done for the 12 $\times$ 12  $\times$ 12 $k$-mesh.
The muffintin (MT) radii are as follows:
$R^{\text{MT}}_{\text{Hg(HgBa2CuO4)}}=$ 2.6 bohr, $R^{\text{MT}}_{\text{Ba(HgBa2CuO4)}}=$ 3.6 bohr, 
$R^{\text{MT}}_{\text{Cu(HgBa2CuO4)}}=$ 2.15 bohr, $R^{\text{MT}}_{\text{O1(HgBa2CuO4)}}=$ 1.50 bohr (in CuO$_2$ plane), 
$R^{\text{MT}}_{\text{O2(HgBa2CuO4)}}=$ 1.10 bohr (others),
$R^{\text{MT}}_{\text{La(La2CuO4)}}=$ 2.88 bohr, $R^{\text{MT}}_{\text{Cu(La2CuO4)}}=$ 2.09 bohr,
$R^{\text{MT}}_{\text{O1(La2CuO4)}}=$ 1.40 bohr (in CuO$_2$ plane), $R^{\text{MT}}_{\text{O2(La2CuO4)}}=$ 1.60 bohr (others).
The angular momentum of the atomic orbitals is taken into account up to
$l=4$ for all the atoms.

The cRPA and GW calculations use a mixed basis consisting of products of two atomic orbitals and interstitial plane waves~\cite{schilfgaarde06}.
In the cRPA and GW calculation, the 6 $\times$ 6 $\times$ 3 $k$-mesh is employed.
By comparing the calculations with the smaller $k$-mesh, we checked that these conditions give well converged results.
We include bands in [$-26.4$: $122.7$] eV (193 bands) for calculation of the screened interaction and the self-energy.
For entangled bands, we disentangle the target bands from the global Kohn-Sham bands~\cite{miyake09}. 

We expect that the difference arising from the choice of basis functions (for instance, plane wave basis or localized basis) in the DFT calculation is small as was shown in a previous work~\cite{miyake10}.
The $3d$ orbital of Cu is relatively localized among that of the transition metals.
Therefore, the bare Coulomb interaction $v$ and the screened Coulomb interaction calculated from $v$ are sensitive to the accuracy of the wave function near the core.
When calculating with a plane wave basis, we would improve the accuracy of interaction by using hard pseudo potentials.

\subsubsection{Method and Conditions for VMC}
We use the open-source software {mVMC~\cite{TaharaVMC,misawaHubbard,mVMC_CPC,mVMC}} that implements the VMC with the variational wave function defined as
\begin{eqnarray}
|\psi\ra =\sP_{\rm G}\sP_{\rm J}\sL^{S}|\phi_{\rm pair}\ra,
\label{mvmc_wavefunction_def}
\end{eqnarray}
where $\sP_{\rm G}$ and $\sP_{\rm J}$
are the Gutzwiller factor~\cite{Gutzwiller} 
and the Jastrow factor~\cite{Jastrow}, respectively.
The variational wave function $|\psi\ra$ is capable of describing
various phases such as magnetic, superconducting, and spin liquid phases in a unified fashion.
We employ the total spin projection $\sL^{S}$
to restore the symmetry of the Hamiltonian~\cite{ring2004nuclear}.
In most part of the calculations,
we use spin singlet total spin projections ($S=0$), which is expected to be the ground-state quantum number.
The pair-product part $|\phi_{\rm pair}\ra$ is
the generalized pairing wave function defined as
\begin{eqnarray}
|\phi_{\rm pair}\ra= \Big[\tc{\sum_{i,j=1}^{\Ns}\sum_{\nu,\mu=1}^{N_{\rm orb}}}f_{ij\nu\mu}\tc{d_{i\nu\uparrow}^{\dag}d_{j\mu\downarrow}^{\dag}}\Big]^{N_{e}/2} |0 \ra,
\end{eqnarray}
where $f_{ij}$ denotes the variational parameters,
$N_{\rm orb}$ is the number of the orbitals, and $N_{\rm s}$ is the number of the lattice sites.
In this calculations, 
we take $2\times2$ sublattice shown in Fig.~\ref{dpp2x2} to consider off-site correlations.
The translational symmetry is assumed beyond this supercell. We have $2\times2\times N_{\rm orb}^2 \times N_{\rm s}$
independent variational parameters for {pair-product} part.
All the variational parameters are simultaneously
optimized by using the stochastic
reconfiguration method~\cite{Sorella_PRB2001,TaharaVMC}.

\section{Result}

\subsection{HgBa$_2$CuO$_4$}

\subsubsection{Band structure in the GWA}

\begin{figure}[h]
\centering 
\includegraphics[clip,width=0.4\textwidth ]{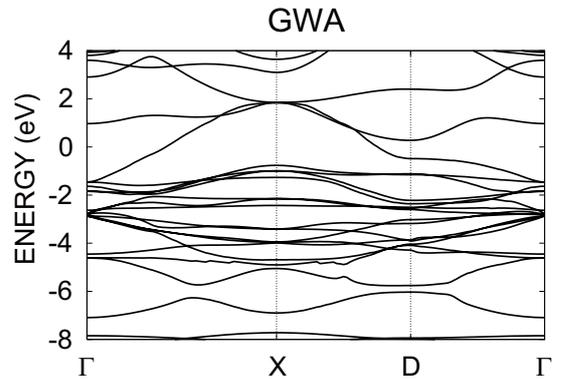} 
\caption{(Color online) 
Electronic band structures of HgBa$_2$CuO$_4$ obtained by the GWA.
Self-energy is calculated only for 
the 17 bands originating from the Cu$3d$ and O $2p$ orbitals near the Fermi level 
indicated by red (gray in black and white plot) bands.
The zero energy corresponds to the Fermi level. 
}
\label{bndHgGW}
\end{figure} 

We show the band structure of HgBa$_2$CuO$_4$ obtained by the GWA in Fig.~\ref{bndHgGW}.
The 17 Wannier functions are constructed from the 20 bands near the Fermi level~\cite{hirayama18}.
Full GW self-energy is introduced to the 17 bands near the Fermi level originating from the Cu $3d$ and O $2p$ orbitals, which are relatively well isolated from higher-energy bands. 
The Cu $3d$ orbitals are split into $t_{2g}$ and $e_g$ orbitals by the octahedral crystal field,
and the $e_g$ orbitals are further split into higher $x^2-y^2$ and lower $3z^2-r^2$ by the crystal field mainly from
the distorted octahedron of the oxygen ions surrounding the copper ions.
The Cu $e_g$ and O $2p$ orbitals are strongly hybridized with each other and make a covalent bond.
Especially, the Cu $x^2-y^2$ orbital has a strong $\sigma $-bonding with the O $2p_{\sigma}$ orbitals directed to the Cu atom,
which makes large band width {$\sim 3.5$ eV}. 
The $s$-band originating from the Hg atom is also hybridized with 17 bands near the Femi level.
The one-body Hamiltonian parameters at the level of full GWA is listed in Table~\ref{paraHg3_cGW-SIC}.

\subsubsection{Three-band Hamiltonian in the cGW-SIC}

\begin{figure}[h]
\centering 
\includegraphics[clip,width=0.4\textwidth ]{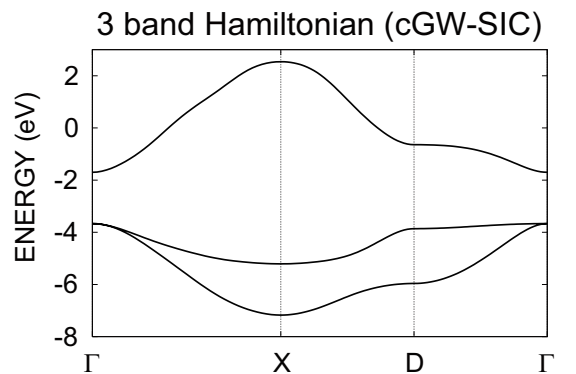} 
\caption{(Color online) Electronic band structure of three-band Hamiltonian in the cGW-SIC originating from the Cu $d_{x^2-y^2}$ and O $2p$ Wannier orbitals for HgBa$_2$CuO$_4$.
The zero energy corresponds to the Fermi level. 
}
\label{bndHgGWcGWSIC3}
\end{figure} 

The three-band Hamiltonian consists of the Cu $d_{x^2-y^2}$ and O $2p_{\sigma}$ orbitals.
The energy window for the Wannier functions is set as the same as that in the GWA for the 17 bands. 
Band structure of the one-body part of three-band {Hamiltonian} in the cGW-SIC is shown in Fig.~\ref{bndHgGWcGWSIC3}.
In the calculation of the cGW-SIC and the cRPA, we use the Green's function obtained by the GWA.
The difference in the chemical potential between the Cu $x^2-y^2$ and O $2p$ orbitals is 2.42 eV.
The nearest-neighbor hopping between the Cu $x^2-y^2$ and O $2p$ orbitals is calculated to be
1.26 eV, which makes a strong covalent bond.
The on-site interaction for the Cu $x^2-y^2$ orbital is strong (8.84 eV), while that in the O $2p$ orbital is relatively weak (5.31 eV).
Details are discussed in Ref.~\onlinecite{hirayama18} and the Hamiltonian parameters are reproduced in Table~\ref{paraHg3_cGW-SIC}.

\subsubsection{Chemical potential correction for three-band Hamiltonian by VMC}

\begin{figure}[h]
\centering 
\includegraphics[clip,width=0.3\textwidth ]{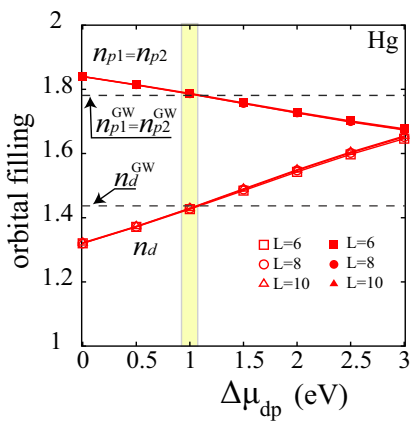} 
\caption{(Color online) 
$\Delta\mu_{dp}$ dependence of the orbital fillings $n_{\nu}$ for the mother material of Hg compound
calculated by VMC for the Hamiltonian (\ref{Hamiltonian0}) with the system size $L\times L$.
Dashed lines show the orbital fillings obtained by the GW calculation.
The proper chemical potential shift is estimated as $\Delta\mu_{dp}=1.0$ eV 
for the Hg compound.
}
\label{occHgVMC3}
\end{figure} 

By using the mVMC, we here analyze the three-band ($dp$) Hamiltonian of HgBa$_2$CuO$_4$ obtained above by the cGW-SIC in the form of Eq.(\ref{Hamiltonian0})\cite{hirayama18}.
The matrix elements of the Hamiltonian
are listed in Table \ref{paraHg3_cGW-SIC}, but the relative level difference between the Cu $d_{x^2-y^2}$ and O $2p_{\sigma}$ Wannier orbitals,
$\Delta\mu_{dp}=\Delta \mu_p - \Delta \mu_d$, is added to tune the orbital fillings.

\begin{figure}[h]
\centering 
\includegraphics[clip,width=0.45\textwidth ]{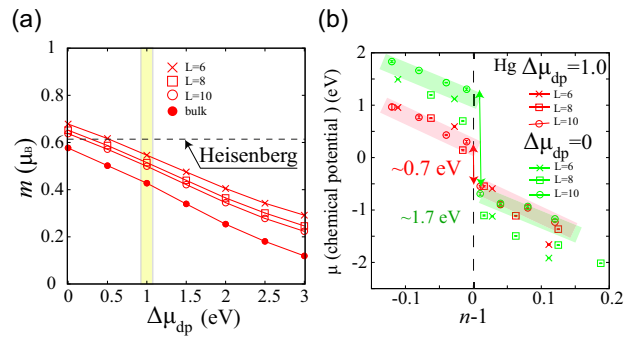} 
\caption{(Color online) 
mVMC results for the Hg three-band Hamiltonian (\ref{Hamiltonian0}) 
with the system size $L\times L$; (a) $\Delta\mu_{dp}$ dependence of the magnetic ordered moment.
{By performing extrapolation with the least square fitting 
for magnetic ordered moments in finite system sizes,
we obtain the bulk (thermodynamic) limit of the magnetic ordered moment. 
At the appropriate correction ($\Delta\mu_{dp}\sim 1$eV), we obtain $m\sim0.4$ ($\mu_{\rm B}$).
For comparison, we show the magnetic ordered moment 
in the Heisenberg model on the square lattice~\cite{Sandvik} by the dashed line.}
(b) Doping dependence of the chemical potential $\mu$.
At the appropriate corrections,
The charge gap is estimated as $\Delta_{c}\sim0.7$ eV.
}
\label{eneHgVMC3}
\end{figure} 

Figure~\ref{occHgVMC3} shows the orbital fillings of the Cu $d_{x^2-y^2}$ and O $2p_{\sigma}$ orbitals as a function of $\Delta\mu_{dp}$ added to the cGW-SIC {Hamiltonian}.
Here, we note that $\Delta\mu_{dp}=0$ corresponds to the cGW-SIC Hamiltonian used in the previous studies~\cite{Misawa_JPSJ2011,hirayama18}
after eliminating the double counting  in the Hartree terms.
Increasing
$\Delta\mu_{dp}$, in other words, decreasing the level difference,
enhances the hybridization between the Cu $d_{x^2-y^2}$ and O $2p_{\sigma}$ orbitals.
The O $2p_{\sigma}$ orbital component becomes larger in the anti-bonding band crossing the Fermi level,
and the filling of the O $2p_{\sigma}$ Wannier orbital decreases.
By taking $\Delta\mu_{dp}=1.0$ eV, the fillings of the O $2p_{\sigma}$ and Cu $d_{x^2-y^2}$ Wannier orbitals meet the values in the GWA.

By using the correction, we calculated the magnetic ordered moment $m$
for HgBa$_2$CuO$_4$ (Fig.~\ref{eneHgVMC3}(a)), 
\tr{which is defined as
\begin{align*}
m=2\Big[\frac{1}{N_{s}}\sum_{i,j}
\langle \boldsymbol{S}_{i}\cdot\boldsymbol{S}_{j}
\rangle e^{i\bm{Q}\cdot(\bm{r}_{i}-\bm{r}_{j})}\Big]^{\frac{1}{2}},
\end{align*}
where $\bm{Q}=(\pi,\pi)$ is the ordering vector.}
The cGW-SIC Hamiltonian without $\Delta \mu$ ($\Delta\mu_{dp}=0$) shows the magnetic ordered moment,
whose amplitude is very close to that of the square lattice Heisenberg model as shown in Fig.~\ref{eneHgVMC3}(a).
Since the existing copper oxide Mott insulators typically show the ordered moment smaller than that of the Heisenberg model~\cite{La214RMP},
the ordered moment is apparently overestimated.
On the other hand, when the correction ($\Delta\mu_{dp}>0$) is taken into account,
the correlation of the system weakens and the magnetic ordered moment is reduced  to a more appropriate value smaller than that in the Heisenberg limit.

The {\it ab initio} matrix elements of
the effective Hamiltonian of the cGW-SIC$+\Delta\mu$ are listed in Table \ref{paraHg3_cGW-SIC}. The difference from cGW-SIC in the same Table is only the level of the $p$ orbital.

By introducing the chemical potential correction, the Mott gap of the effective Hamiltonian at half filling is also estimated using the mVMC.
The Mott gap $\Delta E_{\rm MG}$ is estimated from the total energy difference as $\Delta E_{\rm MG}=(E(N+2)+E(N-2)-2E(N))/2=\mu(N+1)-\mu(N-1)$,
where $E(N)$ and $\mu(N)$ are the ground state energy and the chemical potential of the $N$-electron system, respectively.
Since the Mott gap is formed by strong short-ranged Coulomb repulsion, 
the system size dependence is small and 
the value is a good estimate of the thermodynamic limit.
Here, by introducing the {chemical potential correction
leading to a positive $\Delta\mu_{dp}$,
the hybridization between the $dp$ orbitals becomes stronger and, thus, makes the correlation of the system weaker
than that without the correction.
The Mott gap $\Delta E_{\rm MG}$, indeed, depends on $\Delta\mu_{dp}$: While the Mott gap $\Delta E_{\rm MG}$ without the correction is calculated to be 1.7 eV,
$\Delta E_{\rm MG}$ with the correction $\Delta\mu_{dp}=1$ eV is reduced to 0.7 eV, which proves the weaker correlation in the cGW-SIC$+\Delta \mu$ Hamiltonian,
as shown in Fig.~\ref{eneHgVMC3}(b).
Unfortunately, there exists no mother material in the Hg system.
However, in the next section for the La compound, we will show that our \textit{ab initio} estimate of the Mott gap indeed agrees with the experimental value, in contrast to the estimate without the $\Delta \mu$ correction.

\subsubsection{GW+{LRFB} band}

\begin{figure}[h]
\centering 
\includegraphics[clip,width=0.4\textwidth ]{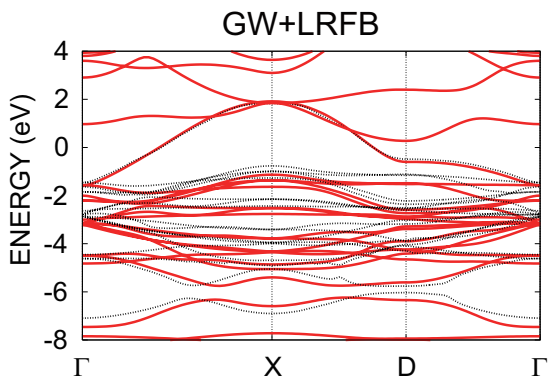} 
\caption{(Color online) 
Electronic band structure of HgBa$_2$CuO$_4$ obtained by the GW{+LRFB} (red solid line).
Self-energy in the GWA is calculated only for the 17 bands originating from the Cu $3d$ and O $2p$ orbitals near the Fermi level.
The feedback effect is counted only to the O $2p$ orbital directed to the Cu atom.
The zero energy corresponds to the Fermi level. 
For comparison, the band structure in the GWA is also given (black dotted line).
}
\label{bndHgGWFB}
\end{figure} 

Next, we calculate the band structure in the GW {combined with LRFB} by adding the on-site correction of the O $2p$ orbitals estimated by the VMC ($\Delta\mu_{dp}=1.0$ eV) to the self-energy in the GWA Green's function.
The chemical potential multiplied by the inverse of the renormalization factor in the cGW-SIC is added to the GW self-energy of the 17 bands near the Fermi level.
We obtain Fig.~\ref{bndHgGWFB} by expanding self-energy to the frequency around the energy eigenvalue of the DFT and diagonalizing the Hamiltonian as the same as that in the usual GWA.
Since there is no frequency dependence in the on-site correction, the frequency dependence of the self-energy remains the same as the GWA, and the renormalization factor is nearly the same as that in the GWA.
The largest change in the GW{+LRFB} band from the GWA is the hybridization between the Cu $x^2-y^2$ and O $2p$ orbitals.
Also, due to the change in the energy level of the O $2p$ orbitals, the 17 bands around the Fermi level is slightly modified through the hybridization.

\subsubsection{Three-band Hamiltonian in cGW-SIC+LRFB}

\begin{figure}[h]
\centering 
\includegraphics[clip,width=0.4\textwidth ]{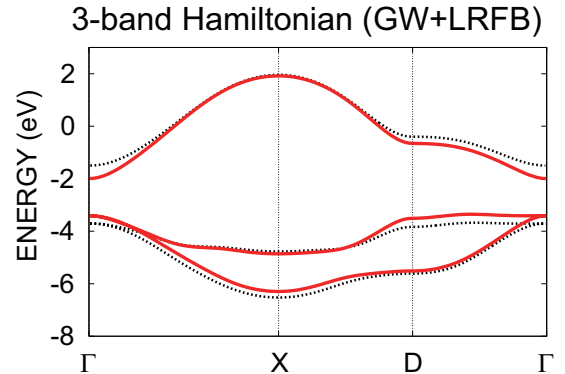} 
\caption{(Color online) Electronic band structure of three-band Hamiltonian in the GW{+LRFB} originating from the Cu $d_{x^2-y^2}$ and O $2p$ Wannier orbitals for HgBa$_2$CuO$_4$.
The zero energy corresponds to the Fermi level. 
For comparison, the band structure of the Wannier function in the original GWA is also given (black dotted line).
}
\label{bndHgGWFBwan3}
\end{figure} 
\begin{figure}[h]
\centering 
\includegraphics[clip,width=0.4\textwidth ]{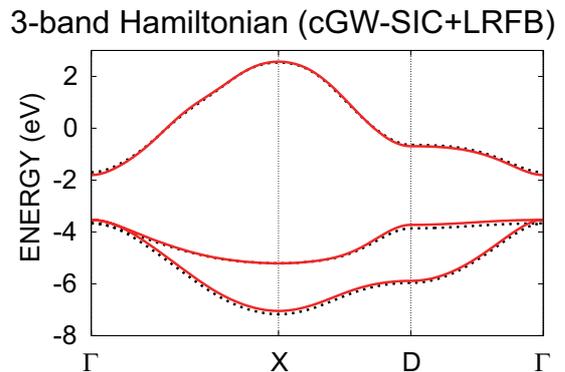} 
\caption{(Color online) Electronic band structure of three-band Hamiltonian consisting of  
the Cu $d_{x^2-y^2}$ and O $2p$ Wannier orbitals for HgBa$_2$CuO$_4$ by using the cGW-SIC+LRFB.
The zero energy corresponds to the Fermi level. 
For comparison, the band structure in the cGW-SIC  
is also plotted (black dotted line).
}
\label{bndHgGWFBcGWSIC3}
\end{figure} 

We derive the 3-band Hamiltonian at the cGW-SIC level based on
the Green's function obtained by
the GW{+LRFB}.
Before deriving cGW-SIC+LRFB, we first show in Fig.~\ref{bndHgGWFBwan3} the band structure
of the 3-band Hamiltonian obtained by
the Wannier function in the level of GW+LRFB.
We set the energy window for the Wannier functions as the same as that in the GWA and GW+LRFB.
Then, the Wannier function of GW+LRFB is close to the atomic orbital similarly to the Wannier function of the GWA.
The bands indicated by the dotted line in the figure is those {obtained by} the Wannier function in the GWA constructed under similar conditions.
The one-body Hamiltonian parameters of GW+LRFB are listed in Table~\ref{paraHg3}.
The chemical potential difference between the Cu $x^2-y^2$ and O $2p$ Wannier orbitals in the GW+LRFB is 1.48 (eV), while that in the GWA is 2.31 (eV) (see Table~\ref{paraHg3_cGW-SIC} for GWA).
The decrease in the chemical potential difference (0.83 eV)
is slightly smaller than $\Delta\mu_{dp}$ (1.0 eV) due to the renormalization factor derived from the static low-energy effective {Hamiltonian}.
Since the Cu $x^2-y^2$ and the O $2p$ orbitals are not hybridized at the $\Gamma$ point due to the symmetry,
the chemical potential change is clearly visible at $\Gamma$ point (Fig.~\ref{bndHgGWFBwan3}).
On the other hand, at the $X$ point, the width of the bonding and anti-bonding bands increases because the energy difference between the Cu $x^2-y^2$ and O $2p$ orbitals decreases.
The correction is a static chemical potential, the $dp$ hopping hardly changes between the GWA (1.18 eV) and the GW{+LRFB} (1.19 eV),
and therefore the increase in the bandwidth of the anti-bonding band is due to purely increase of covalency between the Cu $x^2-y^2$ and O $2p$ orbitals in the GW{+LRFB}.

The band structure obtained by the cGW-SIC+LRFB is shown in Fig.~\ref{bndHgGWFBcGWSIC3}.
The three-band Hamiltonian in the cGW-SIC+LRFB is close to that without the feedback (namely cGW-SIC in Table~\ref{paraHg3_cGW-SIC}).
The Hamiltonian parameters are listed in Table~\ref{paraHg3}. 
The chemical potential difference between the Cu $x^2-y^2$ and the O $2p$ orbitals is 2.17 eV in the cGW-SIC+LRFB, which is close to 2.42 eV in the cGW-SIC obtained from the GW band structure.
Also, the magnitude of the nearest-neighbor hopping between the Cu $x^2-y^2$ and the O $2p$ orbitals is 1.261 eV, which is nearly the same value as 1.257 eV in the cGW-SIC.
The effect of the correction is very small in the three-band Hamiltonian.
This is because the enhanced mutual screening between the Cu $x^2-y^2$ and the O $2p$ orbitals ascribed to the reduced level difference of these two bands is not taken into account at
this stage of the derived three-band Hamiltonian
The screened on-site Coulomb interaction between the Cu $x^2-y^2$ orbitals is 8.99 eV in the cGW-SIC{+LRFB}, while it is 8.84 eV and nearly the same in the cGW-SIC. 
Because of this similarity, the effective three-band Hamiltonian is well represented by cGW-SIC$+\Delta\mu$ when one solves by low-energy solvers.

\begin{figure}[h]
\centering 
\includegraphics[clip,width=0.35\textwidth ]{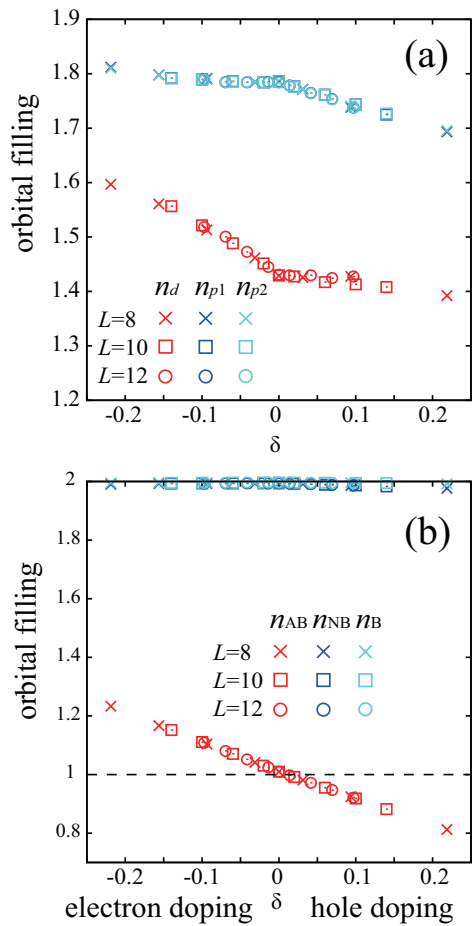} 
\caption{(Color online) 
Doping dependence of the occupation number of the Cu $x^2-y^2$ ($n_d$) and O $2p$ ($n_{p_1}, n_{p_2}$) orbitals for the Hg compound calculated by the VMC for the cGW-SIC$+\Delta\mu$ Hamiltonian. (a) Orbital occupation in the basis of $d_{x^2-y^2}$ and $2p_{\sigma}$ atomic-like Wannier orbitals (b) Orbital occupation in the basis of bonding, nonbonding and antibonding Wannier orbitals, which represents three diagonalized bands in Fig.~\ref{bndHgGWFBcGWSIC3}, respectively.
}
\label{dopeHgVMC3}
\end{figure} 

Figure \ref{dopeHgVMC3} shows the doping concentration ($\delta$) dependence of the orbital fillings for the Cu $3d_{x^2-y^2}$ and two O $2p_{\sigma}$ Wannier orbitals (in (a)) as well as for the Wannier orbitals representing diagonalized bands in Fig.~\ref{bndHgGWFBcGWSIC3} (in (b)) obtained by VMC using the cGW-SIC$+\Delta\mu$ Hamiltonian given in Table~\ref{paraHg3_cGW-SIC}.
Figure~\ref{bndHgGWFBcGWSIC3}(a) shows a kink at zero doping indicating different character of carriers between electron and hole doping.
More remarkably, only the $3d_{x^2-y^2}$ carriers look doped in the electron doped side and only  the $2p$ carriers look doped in the hole doped side around the zero doping, because the filling of the other orbital stays nearly constant,
as was already suggested by the picture of charge transfer insulator~\cite{ZaanenSawatzky}.
This means that
carriers doped in the so-called Hubbard band and lower Hubbard band consist of different orbitals.
However, it is interesting to see the same doping in the Wannier basis functions that represent the bonding, nonbonding and antibonding bands in Fig.~\ref{bndHgGWFBcGWSIC3}, it turns out that the carriers are doped only in the highest antibonding band, as is expected.
This shows that such different characters of carriers in (a) arise only within the carriers belonging to the original anti-bonding band~\cite{anderson}.
Therefore the present apparent charge transfer insulator is well represented by the single-band framework of the antibonding band, which consists of strongly hybridized $d$ and $p$ atomic-like Wannier orbitals.

\subsubsection{Two-band Hamiltonian in cGW-SIC+LRFB}

\begin{figure}[h]
\centering 
\includegraphics[clip,width=0.4\textwidth ]{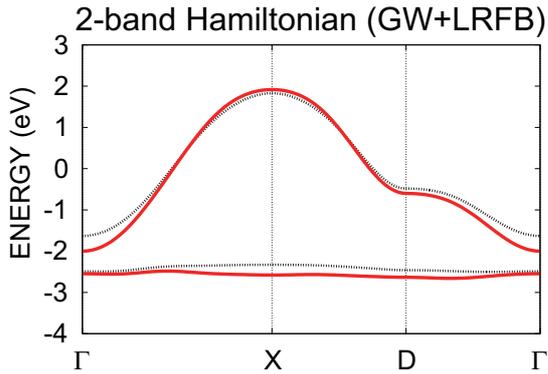} 
\caption{(Color online) Electronic band structure of two-band Hamiltonian in the GW+LRFB originating from the Cu $e_g$ Wannier orbitals for HgBa$_2$CuO$_4$.
The zero energy corresponds to the Fermi level. 
For comparison, the band structure of the Wannier function in the GWA is also given (black dotted line).
}
\label{bndHgGWFBwan2}
\end{figure} 
\begin{figure}[h]
\centering 
\includegraphics[clip,width=0.4\textwidth ]{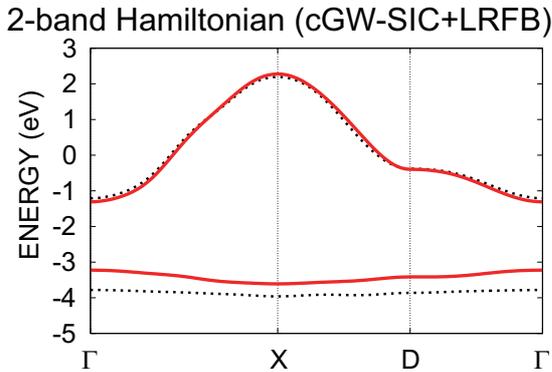} 
\caption{(Color online) Electronic band structure of two-band Hamiltonian 
in the cGW{+LRFB}
originating from the Cu $e_g$ Wannier orbitals for HgBa$_2$CuO$_4$.
The zero energy corresponds to the Fermi level. 
For comparison, the band structure in the cGW-SIC obtained from the GWA is also given (black dotted line).
}
\label{bndHgGWFBcGWSIC2}
\end{figure} 

Next, we discuss the two-band {Hamiltonian} in the cGW-SIC calculated from the GW{+LRFB} band structure (namely cGW-SIC+LRFB band).
The 17 bands around the Fermi level is included to the energy window for the Wannier functions, where the bonding and non-bonding bands of the O $2p$ orbitals are not included. 
The one-body parameters obtained from the full GWA and the cGW-SIC Hamiltonian parameters obtained using the full GW Green's functions are listed in
Table~\ref{paraHg2v0}.  Interaction parameters in the level of cGW-SIC based on the full GW Green's function are calculated by the cRPA and listed in Table~\ref{paraHg2v0} as well. 
Then the level renormalization of O $p_{\sigma}$ orbital is taken into account for the full GW calculation as GW+LRFB in the same way as the three-band calculation.
Figure~\ref{bndHgGWFBwan2} shows the GW+LRFB band structure. The one-body parameters by the GW+LRFB are listed in Table~\ref{paraHg2}. Then the cGW-SIC for the purpose of constructing the two-band (Cu $d_{x^2-y^2}$ and O $p_{\sigma}$) Hamiltonian is performed~\cite{hirayama18}. The one-body parameters for the cGW-SIC+LRFB are listed in Table~\ref{paraHg2}. The interaction parameters for the cGW-SIC+LRFB Hamiltonian are calculated using cRPA applied to the GW+LRFB Green's functions and are listed in Table~\ref{paraHg2} as well.

The energy difference between the anti-bonding orbital and Cu $z^2$ orbital is 2.45 eV in the GW+LRFB,
which is nearly the same as that in the GWA (2.43 eV).
On the other hand, the nearest-neighbor hopping is 0.512 eV in the GW+LRFB, which is factor 1.13 larger than the value of 0.453 eV in the GWA.
This is because the on-site correction increases the contribution of the O $2p$ orbitals to the anti-bonding orbital and then the hopping through the O $2p$ orbital increases.

Band structure in the cGW-SIC+LRFB is shown in Fig.~\ref{bndHgGWFBcGWSIC2}.
The Cu anti-bonding orbitals in the cGW-SIC with the feedback (cGW-SIC+LRFB) is substantially different from that without the feedback (cGW-SIC). 
The effective Hamiltonian parameters are listed in Table~\ref{paraHg2}.
The hybridization amplitude ((nearest-neighbor) transfer integral) of Cu $x^2-y^2$ orbitals with the O $2p$ increases from the cGW-SIC (-0.426eV) to cGW-SIC+LRFB (-0.455 eV), because
the Wannier function of the anti-bonding orbital expands. 
Due to the expansion, the effective interaction decreases.
For instance, the onsite interaction $U$ for the anti-bonding ($d_{x^2-y^2}$) orbital decreases from 4.508 eV (cGW-SIC) to 4.029 eV (cGW-SIC+LRFB). Then the ratio $U/t$ substantially decreases from 10.6  to 8.85.

\subsubsection{One-band {Hamiltonian} in cGW{+LRFB}}

\begin{figure}[h]
\centering 
\includegraphics[clip,width=0.4\textwidth ]{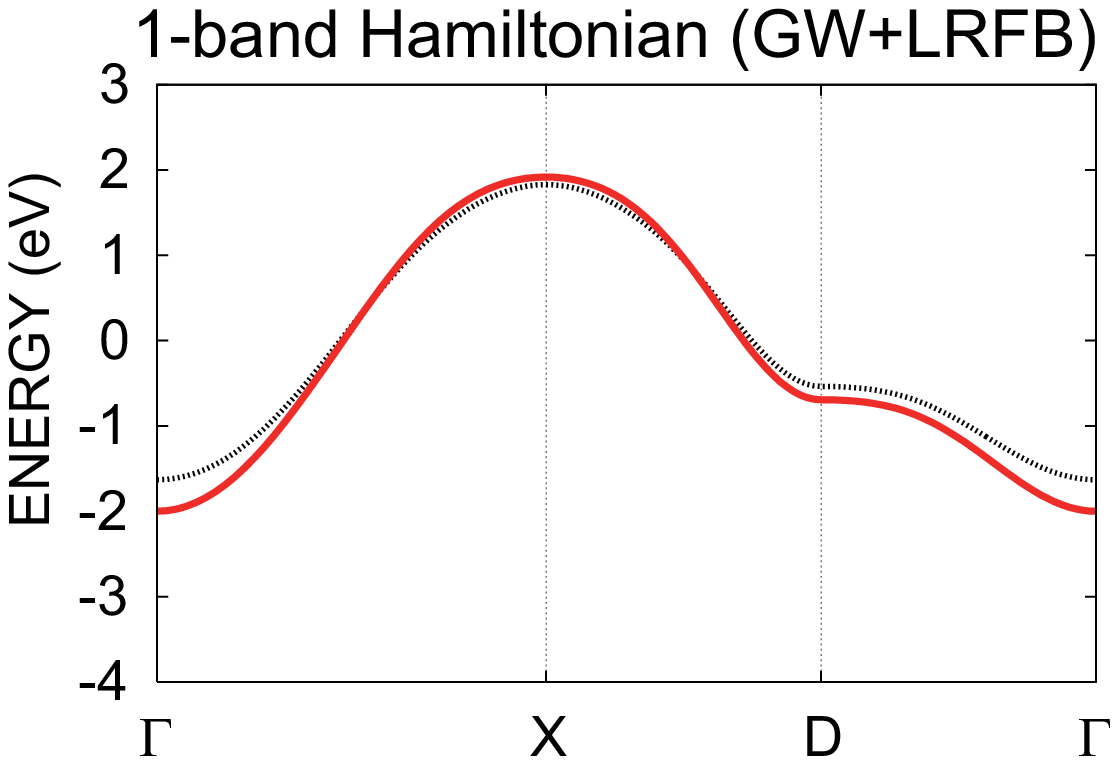} 
\caption{(Color online) Electronic band structure of one-band Hamiltonian in the GW{+LRFB} originating from the Cu $d_{x2-y^2}$ Wannier orbitals for HgBa$_2$CuO$_4$.
The zero energy corresponds to the Fermi level. 
For comparison, the band structure of the Wannier function in the GWA is also given (black dotted line).
}
\label{bndHgGWFBwan1}
\end{figure} 
\begin{figure}[h]
\centering 
\includegraphics[clip,width=0.4\textwidth ]{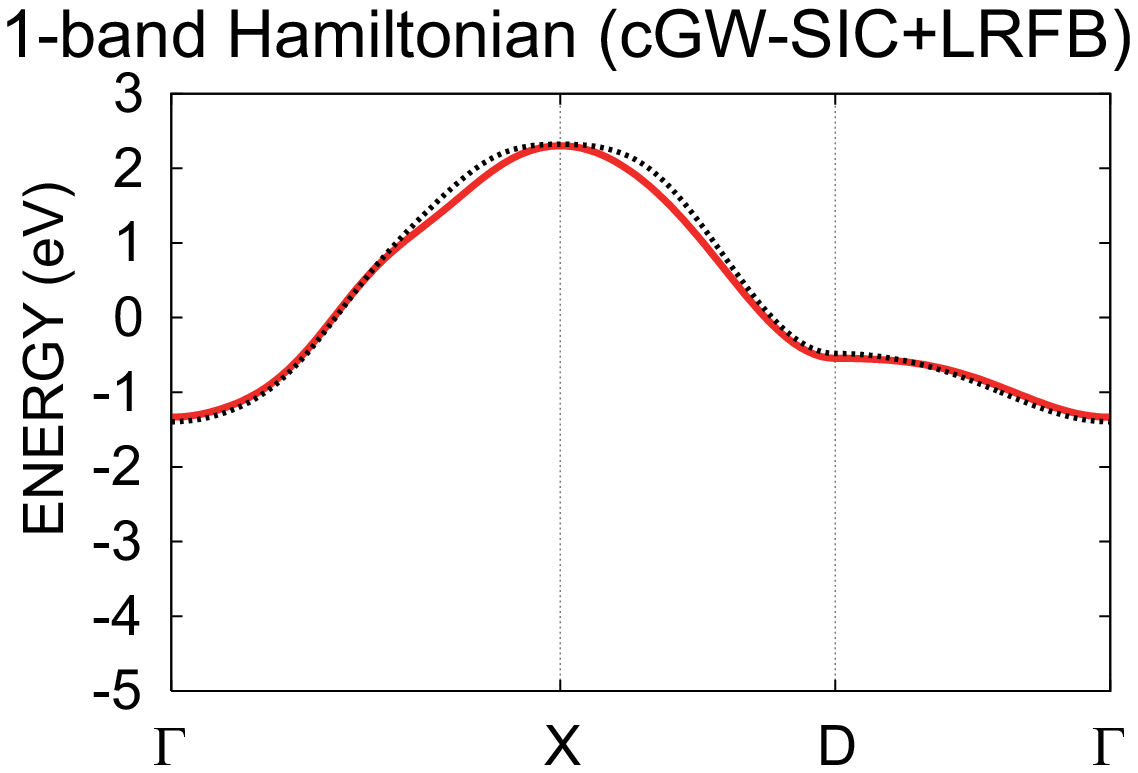} 
\caption{(Color online) Electronic band structure of one-band Hamiltonian in the cGW obtained from the GW{+LRFB} band structure originating from the Cu $d_{x2-y^2}$ Wannier orbitals for HgBa$_2$CuO$_4$.
The zero energy corresponds to the Fermi level. 
For comparison, the band structure in the cGW obtained from the GWA is also given (black dotted line).
}
\label{bndHgGWFBcGW1}
\end{figure} 

The band structure of effective one-band Hamiltonian in the level of GW+LRFB is shown in Fig.~\ref{bndHgGWFBwan1} and the Hamiltonian parameters are listed in Table~\ref{paraHg1v0}.   
The band structure and the one-band Hamiltonian parameters at the level of cGW+LRFB is derived similarly after considering the level correction and feedback, which are shown in Fig.\ref{bndHgGWFBcGW1} and Table~\ref{paraHg1}, respectively. In the case of the one-band Hamiltonian, we do not need to consider the SIC.
The cGW+LRFB Hamiltonian is distinct from the cGW Hamiltonian, where the transfer amplitudes are increased from -0.461 (0.119) eV to -0.509 (0.127) eV for the transfers between the nearest-neighbor sites $t$ (between the next-nearest-neighbor sites $t'$), while the matrix elements of the Coulomb repulsion are decreased from 4.37 (1.09) to 3.85 (0.83) eV for onsite interaction $U$ (nearest-neighbor interaction $V$). The combined effect drives the system into weaker correlation, where  $U/|t|$ is modified from 9.48 to 7.56.

Finally, the parameters for the three types of the effective Hamiltonians for the Hg compounds are summarized in Table~\ref{parameter_summary}.

\subsection{La$_2$CuO$_4$}

\subsubsection{Band structure in the GWA}

The band structure of La$_2$CuO$_4$ based on the GWA is calculated in the same way as the Hg compound and plotted in Fig.~\ref{bndLaGW}.
\begin{figure}[h]
\centering 
\includegraphics[clip,width=0.4\textwidth ]{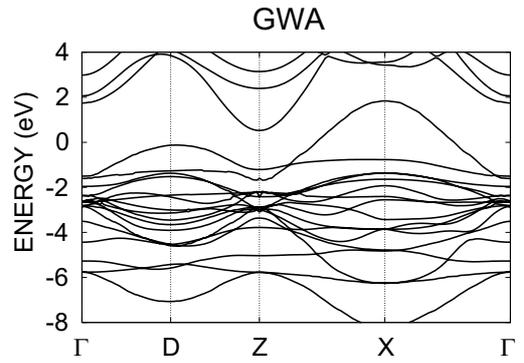} 
\caption{(Color online) 
Electronic band structure of La$_2$CuO$_4$ obtained by the GWA.
Self-energy is calculated for the 17 bands originating from the Cu $3d$ and O $2p$ orbitals near the Fermi level and the $28$ conduction bands originating from the La $4f$ orbitals.
The zero energy corresponds to the Fermi level. 
The 17 bands are drawn in red color (gray in black and white print). 
}
\label{bndLaGW}
\end{figure} 

\subsubsection{\txr{Three}-band Hamiltonian in the cGW-SIC}

The three-band effective Hamiltonian based on the cGW-SIC given in Ref.\onlinecite{hirayama18} is reproduced in Fig.\ref{bndLaGWcGWSIC3}.  

\begin{figure}[h]
\centering 
\includegraphics[clip,width=0.4\textwidth ]{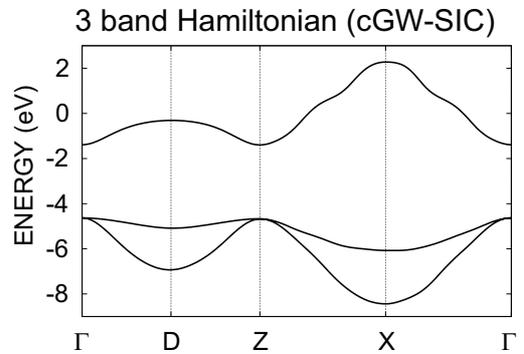} 
\caption{(Color online) Electronic band structure of three-band Hamiltonian in the cGW-SIC originating from the Cu $d_{x^2-y^2}$ and O $2p$ Wannier orbitals for La$_2$CuO$_4$\cite{hirayama18}.
The zero energy corresponds to the Fermi level. 
}
\label{bndLaGWcGWSIC3}
\end{figure} 

\subsubsection{On-site potential correction for three-band Hamiltonian obtained by the VMC: cGW-SIC$+\Delta\mu$}

\begin{figure}[h]
\centering 
\includegraphics[clip,width=0.3\textwidth ]{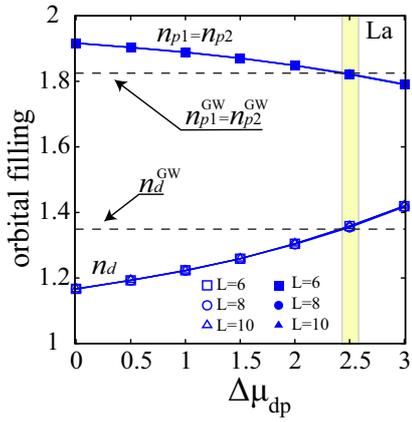} 
\caption{(Color online) 
$\Delta\mu_{dp}$ dependence of the orbital fillings $n_{\nu}$ for the La compound calculated by the VMC at the system size $L\times L$.
Dashed lines show the orbital fillings obtained by the GW calculation.
The appropriate LR (chemical potential correction) is estimated as 2.5 eV.
}
\label{occLaVMC3}
\end{figure} 
\begin{figure}[h]
\centering 
\includegraphics[clip,width=0.45\textwidth ]{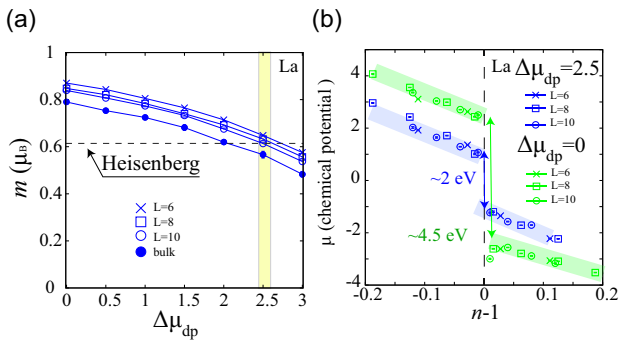} 
\caption{(Color online) 
VMC results of the La compound 
at the system size $L\times L$; (a) $\Delta\mu_{dp}$ dependence of the magnetic ordered moment.
At the proper LR, we obtain $m\sim0.6$ ($\mu_{\rm B}$).
(b) Doping dependence of the chemical potential $\mu$.
At the proper LR,
The charge gap is estimated to be $\Delta_{c}\sim2$eV.
}
\label{eneLaVMC3}
\end{figure} 

\tbb{
Figure~\ref{occLaVMC3} shows $\Delta\mu_{dp}$ dependence of the orbital fillings for La$_2$CuO$_4$.
We find that the proper LR (corrections in the chemical potential) are given by $\Delta\mu_{dp}\sim 2.5$ eV for La$_{2}$CuO$_{4}$.
Therefore we add $\Delta\mu_{dp}\sim 2.5$ eV to the chemical potential of O$p$ orbital.
The revised Hamiltonian parameter on the level of cGW-SIC+$\Delta\mu$ is listed in Table~\ref{paraLa3_cGW-SIC}}

This modified Hamiltonian was solved by mVMC.
The obtained magnetic ordered moments and the chemical potentials are shown in Figs.~\ref{eneLaVMC3}(a) and (b), respectively.
Our calculations show that the magnetic ordered moment for La$_{2}$CuO$_{4}$ is about $0.6$ $\mu_{\rm B}$
\txm{in agreement with the neutron scattering result, $0.60\pm0.05\mu_{\rm B}$~\cite{Yamada}. The charge gap is about $2$ eV, which is consistent with the  available experimental result, for instance the optical conductivity\cite{Uchida}.}

Since the Hamiltonian parameters are remarkably similar for the Hg compounds  between the cGW-SIC and cGW-SIC+LRFB, we also assume that
the parameters for the La compounds estimated by the cGW-SIC+LRFB change little from the cGW-SIC parameters and we do not list here.  Therefore when one solves by using the low-energy solver, the effective three-band Hamiltonian is given just by raising up
the chemical potential of O $p$ orbitals with the amount of 2.5 eV as listed in Table~\ref{paraLa3_cGW-SIC} as cGW-SIC$+\Delta\mu$. The interaction parameters to be used by the low-energy solver are given in the same table, which are obtained by cRPA with the GW-LRFB Green's function.

\begin{figure}[h]
\centering 
\includegraphics[clip,width=0.35\textwidth ]{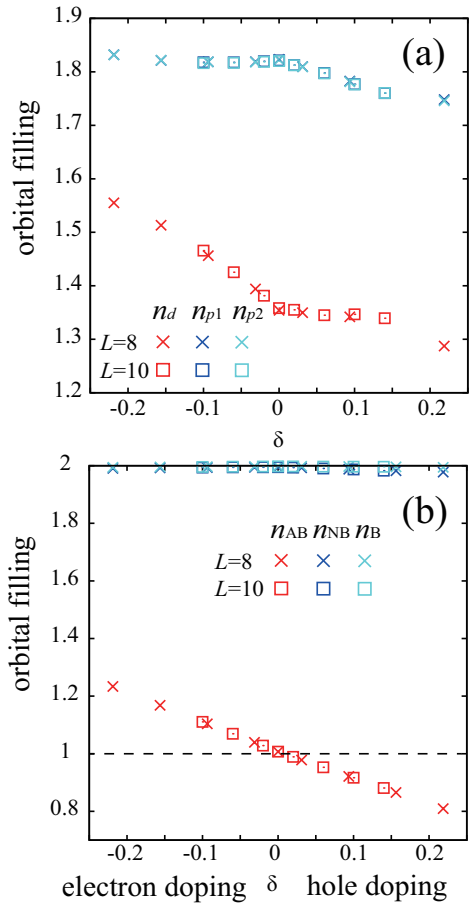} 
\caption{(Color online) 
Doping dependence of the occupation number of the Cu $x^2-y^2$ and O $2p$ orbitals in La calculated by the VMC.
(a) Orbital occupation in the basis of $d_{x^2-y^2}$ and $2p_{\sigma}$ atomic-like Wannier orbitals (b) Orbital occupation in the basis of bonding, nonbonding and antibonding Wannier orbitals, which represents three diagonalized bands in Fig.~\ref{bndLaGWcGWSIC3}, respectively.
}
\label{dopeLaVMC3}
\end{figure} 

Figure \ref{dopeLaVMC3} shows that the carrier character is different between the hole and electron doping in the atomic-like Wannier orbitals in (a) while carriers are solely doped in the antibonding band in (b) similarly to the Hg compound.  This suggests that the two compounds belong to the same class of three-band level scheme, which is essentially described by the single-band framework.

\subsubsection{Strong hybridization of $p$, $d_{3z^2-r^2}$ and $d_{x^2-y^2}$ orbitals in the La compound}
If we try to derive a single-band Hamiltonian in a similar way to the Hg compound, one encounters a difficulty, where strongly hybridizing $d_{3z^2-r^2}$ and the anti-bonding band constructed from the $d_{x^2-y^2}$ and $p_{\sigma}$ orbitals generate substantial off-diagonal self-energy between the $d_{3z^2-r^2}$ and the anti-bonding bands.  However, when we derive the effective one-band Hamiltonian, the entanglement between the anti-bonding and the  $d_{3z^2-r^2}$ orbitals has to be disentangled and the off-diagonal part of the self-energy  has to be ignored.  If only the diagonal self-energy for the anti-bonding band is retained, this truncation results in unphysical wigly behavior of the bands, which is much more serious than the case of the cGW-SIC discussed in Ref.\onlinecite{hirayama18}.  This suggests that the quantitatively precise estimate of the electronic properties must be estimated by including the  $d_{3z^2-r^2}$ orbital degrees of freedom  in the effective Hamiltonian.
Therefore we do not derive the effective single-band Hamiltonian for the La compound.

When we attempt to derive the effective two-band Hamiltonian, the disentanglement and elimination of the non-bonding and anti-bonding bands and resultant neglect of the off-diagonal self-energy involving the bonding/non-bonding electrons again induces weired wavy structure in the two bands, suggesting the necessity to include the bonding/nonbonding states.  Therefore, from the obtained band structure, the reasonable effective Hamiltonian can be obtained only for three-band Hamiltonian or four-band Hamiltonian including all the $e_g$ and $p_{\sigma}$ orbitals on the present level of cGW-SIC+LRFB.

\section{Conclusion and Outlook}

We have derived three-types (three-band, two-band and one-band) of effective Hamiltonians for the HgBa$_2$CuO$_4$ and three-band effective Hamiltonian for La$_2$CuO$_4$ beyond the cGW-SIC effective Hamiltonians derived in Ref.\onlinecite{hirayama18} by improving the treatment of the interband Hartree energy.
More complete effective Hamiltonian parameters including the transfers and interactions at farther distances are listed in Tables in Supplementary Materials.

The necessity of this improvement is clear in our estimates of the Mott gap and antiferromagnetic ordered moment, if one wishes realistic estimates with predictive power. In other words, quantitative accuracy of our derived Hamiltonians by the cGW-SIC+LRFB (or cGW-SIC+$\Delta\mu$) is proven from our VMC solution of the three-band effective Hamiltonian for the La compound: The Mott gap estimated as 2eV and  0.6 $\mu_{\rm B}$ for the antiferromagnetic ordered moment are in good agreement with the experimental results of La$_2$CuO$_4$. 
Although the cuprate compounds have rather complicated band structure with entanglement, the present MACE scheme offers a reasonably accurate effective Hamiltonian for the purpose of understanding physics of copper oxide superconductors.

The obtained Hamiltonians will further serve to clarify physical properties of these copper oxide superconductors, particularly for carrier doped cases, where the mechanism of high-$T_c$ superconductivity remains to be a grand challenge in condensed matter physics. We will discuss physics and properties of carrier doped cases including superconducting properties in a separate publication.

\appendix

\section{Rigidity of orbital filling}
\label{appendix_rigidity}
To examine the rigidity of the orbital occupations, we estimate the energy cost to change the orbital occupation
by employing the following simple charge diagonal part of Coulomb energy,  
\begin{eqnarray}
&&V_{\rm C} = \frac{v_{d}}{2}n_{d}^2 +2v'_{d}n_d^2+\frac{1}{2}v_{p}(n_{p_1}^2+n_{p_2}^2)+v'_pn_{p_1}n_{p_2} \nonumber \\
&+& v_{d,p} n_d (n_{p_1}+n_{p_2})
- \mu_d n_d - \mu_p (n_{p_1}+n_{p_2}),
\end{eqnarray}
where the bare intra-orbital onsite Coulomb interaction between two electrons at the $d_{x^2-y^2} (2p_{\sigma})$ orbital is denoted by $v_{d} (v_{p}) $ and the bare onsite inter-orbital Coulomb interaction between electrons at the $d_{x^2-y^2}$ and $2p_{\sigma}$ is $v_{d,p}$ and $v'_{d}$ and $v'_{p}$ are nearest-neighbor intra-orbital interaction of the $d_{x^2-y^2}$ and $2p_{\sigma}$ orbitals, respectively.  Here, we take into account only up to the nearest-neighbor interaction, because they are the dominant terms.

In this analysis, we only take into account the atomic Coulomb repulsions of the Cu $x^2-y^2$ and O $2p$, because the interaction between a $d_{x^2-y^2}$ or $2p$ electron and an electron at other orbitals are considered in the chemical potential $\mu_d$ and $\mu_p$ provided that the levels of other orbitals are far from the Fermi level and their fillings are rigidly full or empty.
Of course, in $\mu_d$ and $\mu_p$, the potential from the nuclei is also included.

Under the constraint $n_d+2n_p=5$, and $n_p=n_{p1}=n_{p2}$, the Coulomb energy is rewritten as a function of $n_d$ only, as
\begin{equation}
V_{\rm C}=An_d^2+Bn_d+C,
\end{equation}
where $A$ and $B$ are
\begin{eqnarray}
A&=&\frac{v_d}{2}+2v_d'+\frac{v_p+2v_p'}{4}-v_{d,p}, \\
B&=&\frac{5(v_p+2v_p')}{2}+5v_{d,p}-\mu_d+\mu_p
\end{eqnarray}
and $C$ is a constant.
By taking $\delta n_d=n_d-N_d$,
one obtains 
\begin{equation}
V_{\rm C}=A\delta n_d^2+(2N_d+B)\delta n_d+C',
\end{equation}
where $C'=AN_d^2+BN_d+C$. Therefore,when $V_{\rm C}$ has the minimum at $N_d$, the coefficient of the linear term, $2N_d+B=0$ is required.
Then 
\begin{equation}
V_{\rm C}=A(n_d-N_d)^2+C',
\label{eq:VC3}
\end{equation}
is obtained.
When the relative filling between $n_d$ and $n_p$ changes, the energy cost is given by Eq.(\ref{eq:VC3}).

Suppose this interaction energy gives the minimum at $N_d=1.437$ as it is estimated by the full GW calculation (see Table~\ref{paraHg3_cGW-SIC} ).
The effect of strong correlation on the orbital occupation beyond the GW approximation can be roughly estimated from the solution of the mVMC within the effective three-band Hamiltonian of Hg compound with the parameters listed in Table~\ref{paraHg3_cGW-SIC}.  The mVMC energy for several choices of lattice sizes is plotted in Fig.~\ref{fig:E_VMC}. Since the size dependence is small, we employ $L=10$ result, as the thermodynamic limit. Strong correlation effects makes the $d$-orbital filling smaller from the GW value, $n_d=$1.437 to 1.32.
\begin{figure}[h!]
\centering 
\includegraphics[width=0.35\textwidth ]{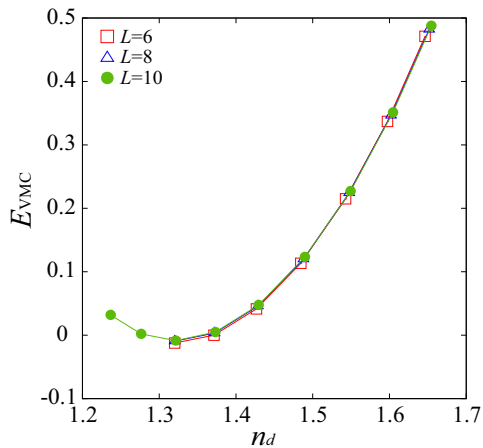} 
\caption{(Color online) 
$n_d$ dependence of total energy per site of the three-band effective Hamiltonian for the Hg compound estimated by the mVMC calculation. The obtained energy has the minimum at $n_d=1.32$.
Here, we added a constant in the energy so as to make the minimum energy zero. 
}
\label{fig:E_VMC}
\end{figure} 

Then although it is not a rigorous treatment, the rigidity of the orbital occupation is roughly estimated by adding the VMC energy $E_{\rm VMC}$ to the bare Coulomb energy given by Eq.(\ref{eq:VC3}) with $A\simeq 20.9$ eV  as can be estimated in the present paper for HgBa$_2$CuO$_4$ (see Table~\ref{paraHg3_cGW-SIC} ). 
This means that electrons in the low-energy degrees of freedom follows the low-energy effective Hamiltonian under the parabolic potential given by Eq.(\ref{eq:VC3}).
Namely, the rigidity of the orbital occupation is roughly estimated by the shift of the minimum from $N_d$ when we add the energy calculated from the solution of the low-energy effective Hamiltonian defined before the level renormalization.

The {\it ab initio} three-band effective Hamiltonian for the Hg compound with the parameters listed in Table~\ref{paraLa3_cGW-SIC} for cGW-SIC 
was solved by the mVMC. The resultant energy $E_{\rm VMC}$ is plotted in Fig.~\ref{fig:Energy_nd}. 
When we plot $V_{\rm C}+E_{\rm VMC}$, $n_d$ which gives the minimum value shifts from the minimum of $V_{\rm C}$ $n_d=1.437$ to 1.415 with the amount 0.022 as one sees in Fig.~\ref{fig:Energy_nd}. 
This little change proves the rigidity of the orbital filling $n_d$ estimated by the GW approximation and justifies the present treatment to fix the orbital occupation determined from the DFT or GW approximation.

The self-consistent dynamical mean-field treatment was formulated by taking account of correlation-induced changes to the total charge density to impose the self-consistency for the charge density\cite{Pourovskii}. It was applied to thin films of SrVO$_3$ and the self-consistent GW treatment shows that the orbital occupation of $d_{xy}$ and $d_{yz}/d_{zx}$ orbitals recovers to values similar to the DFT estimates.~\cite{Bhandary,Schuler}. This again endorses the rigidity of the orbital occupation.

\begin{figure}[h!]
\centering 
\includegraphics[width=0.35\textwidth ]{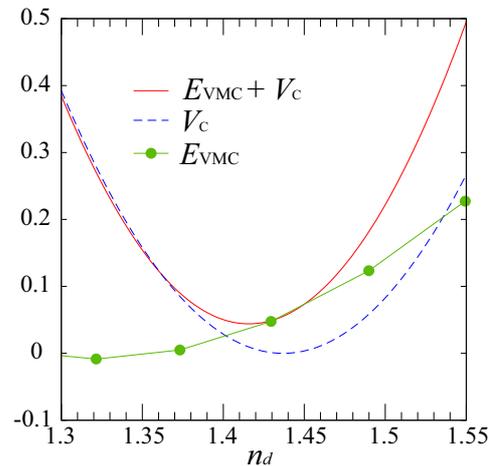} 
\caption{(Color online) 
Comparison of $n_d$ dependence of Coulomb energy $V_{\rm C}$, $E_{\rm VMC}$ and $V_{\rm C}+E_{\rm VMC}$.
Here, the minimum of 
$V_{\rm C}$ is taken to be zero by adding a constant to $V_{\rm C}$.
}
\label{fig:Energy_nd}
\end{figure}

\clearpage

\begin{table*}[h] 
\caption{
Transfer integrals and effective interactions for three-band Hamiltonian of HgBa$_2$CuO$_4$ 
(in eV).
We show the transfer integral in the cGW-SIC$+\Delta\mu$ as well as that in the GWA  for comparison, while the effective interaction is the result of the cRPA. The data for the cGW-SIC$+\Delta\mu$ except for the orbital level of $p_1$ and $p_2$ are taken from Table IV of Ref.\onlinecite{hirayama18} obtained by the cGW-SIC. For the $p_1$ and $p_2$  levels, $\Delta_{dp}=1$ eV is added to the level in Ref.\onlinecite{hirayama18}.
$v$ and $J_{v}$ represent the bare Coulomb and exchange interactions, respectively. $U(0)$ and $J(0)$ represent the static values of the effective Coulomb and exchange interactions, respectively (at $\omega=0$). The index 'n' and 'nn' represent the nearest, [1,0,0] and the next-nearest sites [1,1,0] respectively. 
The occupation number in the GWA is also given in the bottom column ``occu.(GWA)'' in this Table.
The parameters for further neighbor transfer integrals and interactions by the cGW-SIC+$\Delta\mu$ are given in the Supplemental Material~\cite{SM}.
}
\ 
\label{paraHg3_cGW-SIC} 
\begin{tabular}{c|ccc|ccc|ccc|ccc|ccc} 
\hline \hline \\ [-8pt]
$t $(GWA)   &       &  $(0,0,0)$  &       &     & $(1,0,0)$ &    &       & $(1,1,0)$ &      &     &  $(2,0,0)$ &     \\ [+1pt]
\hline \\ [-8pt]
      &  $x^2-y^2$ &  $p_1$ &  $p_2$ & $x^2-y^2$ &  $p_1$ &  $p_2$ &  $x^2-y^2$ &  $p_1$ &  $p_2$ & $x^2-y^2$ &  $p_1$ &  $p_2$ \\ 
\hline \\ [-8pt] 
$x^2-y^2$  & -1.597 & -1.184 &  1.184   & -0.014 & -0.026 & -0.016 &  0.020 & 0.004 & -0.004 & 0.002 & -0.005 & -0.002 \\ 
$p_1$          & -1.184 & -3.909 & -0.659  &   1.184 &  0.111 &  0.659  & -0.016 & 0.039 &  0.003 & 0.026 & -0.008 &  0.003 \\
$p_2$          &  1.184 & -0.659 &  -3.909  &  -0.016 &-0.003 & -0.061 &  0.016 & 0.003 &  0.039  & -0.002 & 0.006 & -0.004    \\
\hline \hline \\ [-8pt]
$t $(cGW-SIC)   &       &  $(0,0,0)$  &       &     & $(1,0,0)$ &    &       & $(1,1,0)$ &      &     &  $(2,0,0)$ &     \\ [+1pt]
\hline \\ [-8pt]
      &  $x^2-y^2$ &  $p_1$ &  $p_2$ & $x^2-y^2$ &  $p_1$ &  $p_2$ &  $x^2-y^2$ &  $p_1$ &  $p_2$ & $x^2-y^2$ &  $p_1$ &  $p_2$ \\ 
\hline \\ [-8pt] 
$x^2-y^2$  &  -1.696 & -1.257 & 1.257  & -0.012 & -0.033 & -0.056 &  0.021 & -0.012  & 0.012 & -0.012 &  0.004 & -0.003 \\
$p_1$          & -1.257 & -4.112 & -0.751 & 1.257  &  0.181  & 0.751  & -0.056 &  0.054  & 0.004 &  0.033 & -0.006 &  0.004 \\
$p_2$          &  1.257 & -0.751 & -4.112 & -0.056 & -0.004 & -0.060  & 0.056 &  0.004  & 0.054 & -0.003 &  0.001 & -0.004 \\
\hline \hline \\ [-8pt]
$t $(cGW-SIC+$\Delta\mu$)   &       &  $(0,0,0)$  &       &     & $(1,0,0)$ &    &       & $(1,1,0)$ &      &     &  $(2,0,0)$ &     \\ [+1pt]
\hline \\ [-8pt]
      &  $x^2-y^2$ &  $p_1$ &  $p_2$ & $x^2-y^2$ &  $p_1$ &  $p_2$ &  $x^2-y^2$ &  $p_1$ &  $p_2$ & $x^2-y^2$ &  $p_1$ &  $p_2$ \\ 
\hline \\ [-8pt] 
$x^2-y^2$  &  -1.696 & -1.257 & 1.257  & -0.012 & -0.033 & -0.056 &  0.021 & -0.012  & 0.012 & -0.012 &  0.004 & -0.003 \\
$p_1$          & -1.257 & -3.112 & -0.751 & 1.257  &  0.181  & 0.751  & -0.056 &  0.054  & 0.004 &  0.033 & -0.006 &  0.004 \\
$p_2$          &  1.257 & -0.751 & -3.112 & -0.056 & -0.004 & -0.060  & 0.056 &  0.004  & 0.054 & -0.003 &  0.001 & -0.004 \\
\hline \hline \\ [-8pt]  
   &       &  $v$  &       &     & $U(0)$ &    &       & $J_{v}$ &      &     &  $J(0)$ &     \\ [+1pt]
\hline \\ [-8pt]
      &  $x^2-y^2$ &  $p_1$ &  $p_2$ & $x^2-y^2$ &  $p_1$ &  $p_2$ &  $x^2-y^2$ &  $p_1$ &  $p_2$ & $x^2-y^2$ &  $p_1$ &  $p_2$ \\ 
\hline \\ [-8pt] 
$x^2-y^2$ & 28.821  &  8.010 &   8.010   & 8.837 &  1.985 &   1.985  &            &  0.063 &   0.063  &           &   0.048 &  0.048  \\ 
$p_1$         &   8.010  & 17.114 &  5.319  &  1.985  & 5.311  &  1.210  &  0.063  &            &  0.041  & 0.048  &  -          &  0.020 \\
 $p_2$        &   8.010  &  5.319  & 17.114  & 1.985  &  1.210 &   5.311 &  0.063   & 0.041  &            &  0.048  & 0.020  &         \\
\hline \hline \\ [-8pt]  
       &       &  $v_{\text{n}}$ &    &     & $V_{\text{n}}(0)$ &    &       & $v_{\text{nn}}$  &      &     &  $V_{\text{nn}}(0)$ &     \\ [+1pt]
\hline \\ [-8pt] 
      &  $x^2-y^2$ &  $p_1$ &  $p_2$ & $x^2-y^2$ &  $p_1$ &  $p_2$ &  $x^2-y^2$ &  $p_1$ &  $p_2$ & $x^2-y^2$ &  $p_1$ &  $p_2$ \\ 
\hline \\ [-8pt] 
$x^2-y^2$ &   3.798 &  8.010  &  3.339  & 0.804  &  1.985   & 0.650   &  2.706  &  3.339  &  3.339  &  0.380  &  0.545  &  0.544  \\
$p_1$         &   2.577  & 3.877  &  2.417  &  0.499  &  0.847  &  0.450  &  2.172  &  2.678  &  2.417  &  0.286  &  0.415  &  0.356 \\
$p_2$        &   3.339  & 5.319  &  3.601   &  0.650  &  1.210  &  0.705  &  2.172  &  2.417  &  2.678  &  0.286  &  0.356  &  0.414 \\
\hline \\ [-8pt]
occ.(GWA)      &  $x^2-y^2$ &  $p_1$ &  $p_2$  \\ 
\hline \\ [-8pt] 
                     & 1.437 & 1.781  & 1.781   \\
\hline
\hline 
\end{tabular} 
\end{table*} 

\begin{table*}[h] 
\caption{
Transfer integrals and effective interactions for three-band Hamiltonian of HgBa$_2$CuO$_4$ (in eV).
Both the transfer integrals and the effective interactions are calculated based on the GW+LRFB band structure.
Transfer integrals denoted by GW+LRFB are obtained from the Wannier orbital constructed to fit the GW+LRFB band structure.  On the other hand,
transfer integrals denoted as cGW-SIC+LRFB
is obtained by the cGW-SIC procedure using the GW+LRFB bandstructure/Green's functions. 
The effective interactions are the result of the cRPA applied to the GW+LRFB Green's functions.
$v$ and $J_{v}$ represent the bare Coulomb and exchange interactions, respectively. $U(0)$ and $J(0)$ represent the static values of the effective Coulomb and exchange interactions, respectively (at $\omega=0$). The index 'n' and 'nn' represent the nearest, [1,0,0] and the next-nearest sites [1,1,0] respectively. 
The occupation number in the GWA is also given in the bottom column ``occ.(GWA)'' in this Table.
The parameters for further neighbor transfer integrals and interactions by the cGW-SIC+LRFB are given in the Supplemental Material~\cite{SM}.
}
\ 
\label{paraHg3} 
\begin{tabular}{c|ccc|ccc|ccc|ccc|ccc} 
\hline \hline \\ [-8pt]
$t $(\tbb{GW+LRFB})   &       &  $(0,0,0)$  &       &     & $(1,0,0)$ &    &       & $(1,1,0)$ &      &     &  $(2,0,0)$ &     \\ [+1pt]
\hline \\ [-8pt]
      &  $x^2-y^2$ &  $p_1$ &  $p_2$ & $x^2-y^2$ &  $p_1$ &  $p_2$ &  $x^2-y^2$ &  $p_1$ &  $p_2$ & $x^2-y^2$ &  $p_1$ &  $p_2$ \\ 
\hline \\ [-8pt] 
$x^2-y^2$  & -2.105 & -1.189 &  1.189   & -0.014 & -0.027  & -0.013 &  0.020 & 0.006 & -0.006 & 0.002 & -0.006 & -0.003 \\ 
$p_1$          & -1.189 & -3.587 & -0.680  &   1.189 &  0.108 &  0.680  & -0.013 & 0.034 &  0.006 & 0.027 & -0.013 &  0.006 \\
$p_2$          &  1.189 & -0.680 &  -3.587  &  -0.013 &-0.006 & -0.063 &  0.013 & 0.006 &  0.0394 & -0.003 & 0.006 & -0.005    \\
\hline \hline \\ [-8pt]
$t $(cGW-SIC+LRFB)   &       &  $(0,0,0)$  &       &     & $(1,0,0)$ &    &       & $(1,1,0)$ &      &     &  $(2,0,0)$ &     \\ [+1pt]
\hline \\ [-8pt]
      &  $x^2-y^2$ &  $p_1$ &  $p_2$ & $x^2-y^2$ &  $p_1$ &  $p_2$ &  $x^2-y^2$ &  $p_1$ &  $p_2$ & $x^2-y^2$ &  $p_1$ &  $p_2$ \\ 
\hline \\ [-8pt] 
$x^2-y^2$  &  -1.801 & -1.261 &  1.261  & -0.014 & -0.034 & -0.056 &  0.023 & -0.011  & 0.012 & -0.012 &  0.004 & -0.004 \\
$p_1$          & -1.261 & -3.975 & -0.753 & 1.261  &  0.183  & 0.753  &  -0.057 &   0.053  & 0.007 &  0.034 & -0.009 &  0.007 \\
$p_2$          &  1.261 & -0.753 & -3.975 & -0.057 & -0.007 & -0.060  & 0.056 &  0.007  & 0.053 &  -0.004 &  0.001 & -0.006 \\
\hline \hline \\ [-8pt]  
   &       &  $v$  &       &     & $U(0)$ &    &       & $J_{v}$ &      &     &  $J(0)$ &     \\ [+1pt]
\hline \\ [-8pt]
      &  $x^2-y^2$ &  $p_1$ &  $p_2$ & $x^2-y^2$ &  $p_1$ &  $p_2$ &  $x^2-y^2$ &  $p_1$ &  $p_2$ & $x^2-y^2$ &  $p_1$ &  $p_2$ \\ 
\hline \\ [-8pt] 
$x^2-y^2$ & 28.821  &  8.010 &   8.010   & 8.986 &  2.053 &   2.053 &            &  0.063 &   0.063  &           &   0.048 &  0.048  \\ 
$p_1$         &   8.010  & 17.114 &  5.319  &  2.053  & 5.404  &  1.253  &  0.063  &            &  0.041  & 0.048  &            &  0.020 \\
 $p_2$        &   8.010  &  5.319  & 17.114  & 2.053  &  1.253 &   5.404 &  0.063   & 0.041  &            &  0.048  & 0.020  &         \\
\hline \hline \\ [-8pt]  
       &       &  $v_{\text{n}}$ &    &     & $V_{\text{n}}(0)$ &    &       & $v_{\text{nn}}$  &      &     &  $V_{\text{nn}}(0)$ &     \\ [+1pt]
\hline \\ [-8pt] 
      &  $x^2-y^2$ &  $p_1$ &  $p_2$ & $x^2-y^2$ &  $p_1$ &  $p_2$ &  $x^2-y^2$ &  $p_1$ &  $p_2$ & $x^2-y^2$ &  $p_1$ &  $p_2$ \\ 
\hline \\ [-8pt] 
$x^2-y^2$ &   3.798 &  8.010  &  3.339  & 0.844  &  2.053   & 0.681   &  2.706  &  3.339  &  3.339  &  0.513  &  0.681  &  0.681  \\
$p_1$         &   2.577  & 3.877  &  2.417  &  0.525  &  0.887  &  0.473  &  2.172  &  2.678  &  2.417  &  0.404  &  0.535  &  0.473 \\
$p_2$        &   3.339  & 5.319  &  3.601   &  0.681  &  1.253  &  0.736  &  2.172  &  2.417  &  2.678  &  0.405  &  0.473  &  0.535 \\
\hline \\ [-8pt]
occ.(GWA)      &  $x^2-y^2$ &  $p_1$ &  $p_2$  \\ 
\hline \\ [-8pt] 
                     & 1.437 & 1.781  & 1.781   \\
\hline
\hline 
\end{tabular} 
\end{table*}

\begin{table*}[h] 
\caption{
Transfer integral and effective interaction in two-band Hamiltonian for HgBa$_2$CuO$_4$ (in eV). We show the transfer integral in the cGW-SIC and also in the GWA for comparison.
The transfer integrals denoted as GWA are calculated from the Wannier orbitals constructed to fit the GW band structure. The transfer integrals denoted as cGW-SIC are calculated from the cGW procedure applied to the GWA band structure/Green's functions.
The effective interaction is the result of the cRPA applied to the GWA Green's functions.
$v$ and $J_{v}$ represent the bare Coulomb interaction/exchange interactions respectively. $U(0)$ and $J(0)$ represent the static values of the effective Coulomb interaction/exchange interactions (at $\omega=0$). 
The data are the same as and taken from Table II of Ref.\onlinecite{hirayama18}.
}
\ 
\label{paraHg2v0} 
\begin{tabular}{c|cc|cc|cc|cc|cc} 
\hline \hline \\ [-8pt]
$t $(GWA)   &       &  $(0,0,0)$  &      & $(1,0,0)$ &       & $(1,1,0)$ &      &  $(2,0,0)$     \\ [+1pt]
\hline \\ [-8pt]
      &  $3z^2-r^2 $ &  $x^2-y^2 $  &  $3z^2-r^2 $ &  $x^2-y^2 $ &  $3z^2-r^2 $ &  $x^2-y^2 $ &  $3z^2-r^2 $ &  $x^2-y^2 $ \\ 
\hline \\ [-8pt] 
$3z^2-r^2 $  &  -2.282 & 0.000 &  -0.018 & 0.084 &  -0.006 & 0.000 & -0.003 &  0.010 \\
$x^2-y^2 $  &  0.000  & 0.144 & 0.084  &-0.453 & 0.000   &  0.074 &  0.010 & -0.051 \\
\hline \hline \\ [-8pt]
$t $(cGW-SIC)   &       &  $(0,0,0)$  &      & $(1,0,0)$ &        & $(1,1,0)$ &        &  $(2,0,0)$     \\ [+1pt]
\hline \\ [-8pt]
      &  $3z^2-r^2 $ &  $x^2-y^2 $  &  $3z^2-r^2 $ &  $x^2-y^2 $ &  $3z^2-r^2 $ &  $x^2-y^2 $ &  $3z^2-r^2 $ &  $x^2-y^2 $ \\ 
\hline \\ [-8pt] 
$3z^2-r^2 $  & -3.811 & 0.000 & 0.013 & 0.033  & -0.003 & 0.000 & 0.000 & 0.002 \\
$x^2-y^2 $  & 0.000  & 0.197 & 0.033 & -0.426 & 0.000  & 0.102 & 0.002 & -0.048 \\
\hline \hline \\ [-8pt]  
   &       &  $v$  &      & $U(0)$ &     & $J_{v}$ &       &  $J(0)$      \\ [+1pt]
\hline \\ [-8pt]
       &  $3z^2-r^2 $ &  $x^2-y^2 $  &  $3z^2-r^2 $ &  $x^2-y^2 $ &  $3z^2-r^2 $ &  $x^2-y^2 $ &  $3z^2-r^2 $ &  $x^2-y^2 $ \\ 
\hline \\ [-8pt] 
$3z^2-r^2 $   & 24.348  & 18.672 &  6.922  & 3.998 &             &   0.808   &               &  0.726 \\
$x^2-y^2 $  & 18.672  & 17.421 & 3.998   & 4.508 & 0.808  &                &  0.726   &             \\ 
\hline \hline \\ [-8pt]  
       &       &  $v_{\text{n}}$ &    & $V_{\text{n}}(0)$ &       & $v_{\text{nn}}$  &       &  $V_{\text{nn}}(0)$     \\ [+1pt]
\hline \\ [-8pt] 
      &  $3z^2-r^2 $ &  $x^2-y^2 $  &  $3z^2-r^2 $ &  $x^2-y^2 $ &  $3z^2-r^2 $ &  $x^2-y^2 $ &  $3z^2-r^2 $ &  $x^2-y^2 $ \\ 
\hline \\ [-8pt] 
$3z^2-r^2 $  & 3.669 &  3.922  & 0.764   & 0.833 & 2.657  & 2.696 & 0.486 &  0.502 \\
$x^2-y^2 $  &   3.922 &  4.155 & 0.833  & 0.901&  2.696 & 2.749 & 0.502 & 0.522 \\
\hline \\ [-8pt]
occ.(GWA)      &  $3z^2-r^2 $ &  $x^2-y^2 $  \\ 
\hline \\ [-8pt] 
                     & 1.992  &  1.008   \\
\hline
\hline 
\end{tabular} 
\end{table*} 

\begin{table*}[h] 
\caption{
Transfer integrals and effective interactions for two-band Hamiltonian of HgBa$_2$CuO$_4$ (in eV).
Both the transfer integrals and the effective interactions are calculated based on the GW+LRFB band structure and the GW+LRFB Green's functions.
The transfer integral denoted as GW+LRFB is obtained from the Wannier orbitals constructed to fit this GW+LRFB band structure.  On the other hand, the transfer integrals denoted as cGW-SIC+LRFB is obtained by the cGW-SIC+LRFB procedure, where the cGW-SIC is applied to the GW+LRFB band structure/Green's functions.  
The effective interaction is the result of the cRPA obtained by using the GW+LRFB Green's functions.
$v$ and $J_{v}$ represent the bare Coulomb interaction/exchange interactions respectively. $U(0)$ and $J(0)$ represent the static values of the effective Coulomb interaction/exchange interactions (at $\omega=0$). The index 'n' and 'nn' represent the nearest unit cell [1,0,0] and the next-nearest unit cell [1,1,0] respectively.
The occupation number in the GWA is also given in this Table.
The parameters for further neighbor transfer integrals and interactions by the cGW-SIC+LRFB are given in the Supplemental Material~\cite{SM}.
}
\ 
\label{paraHg2} 
\begin{tabular}{c|cc|cc|cc|cc|cc} 
\hline \hline \\ [-8pt]
$t $(GW+LRFB)   &       &  $(0,0,0)$  &      & $(1,0,0)$ &       & $(1,1,0)$ &      &  $(2,0,0)$     \\ [+1pt]
\hline \\ [-8pt]
      &  $3z^2-r^2 $ &  $x^2-y^2 $  &  $3z^2-r^2 $ &  $x^2-y^2 $ &  $3z^2-r^2 $ &  $x^2-y^2 $ &  $3z^2-r^2 $ &  $x^2-y^2 $ \\ 
\hline \\ [-8pt] 
$3z^2-r^2 $  &  -2.556 & 0.000 &  0.003 & 0.113 &  -0.011 & 0.000 & -0.006 &  0.012 \\
$x^2-y^2 $  &  0.000  & -0.109 & 0.113  & -0.512 & 0.000   &  0.079 &  0.012 & -0.064 \\
\hline \hline \\ [-8pt]
$t $(cGW-SIC+LRFB)   &       &  $(0,0,0)$  &      & $(1,0,0)$ &        & $(1,1,0)$ &        &  $(2,0,0)$     \\ [+1pt]
\hline \\ [-8pt]
      &  $3z^2-r^2 $ &  $x^2-y^2 $  &  $3z^2-r^2 $ &  $x^2-y^2 $ &  $3z^2-r^2 $ &  $x^2-y^2 $ &  $3z^2-r^2 $ &  $x^2-y^2 $ \\ 
\hline \\ [-8pt] 
$3z^2-r^2 $  & -3.518 & 0.000 &  0.002 & 0.029  & -0.003 & 0.000 & -0.003 & 0.004 \\
$x^2-y^2 $  & 0.000  & 0.187 & 0.029 & -0.455 & 0.000  & 0.096 & 0.004 & -0.040 \\
\hline \hline \\ [-8pt]  
   &       &  $v$  &      & $U(0)$ &     & $J_{v}$ &       &  $J(0)$      \\ [+1pt]
\hline \\ [-8pt]
       &  $3z^2-r^2 $ &  $x^2-y^2 $  &  $3z^2-r^2 $ &  $x^2-y^2 $ &  $3z^2-r^2 $ &  $x^2-y^2 $ &  $3z^2-r^2 $ &  $x^2-y^2 $ \\ 
\hline \\ [-8pt] 
$3z^2-r^2 $   & 21.816  & 17.022 &  5.962  & 3.497 &             &   0.737   &               &  0.645 \\
$x^2-y^2 $  & 17.022  & 16.197 & 3.497   & 4.029 & 0.737  &                &  0.645   &             \\ 
\hline \hline \\ [-8pt]  
       &       &  $v_{\text{n}}$ &    & $V_{\text{n}}(0)$ &       & $v_{\text{nn}}$  &       &  $V_{\text{nn}}(0)$     \\ [+1pt]
\hline \\ [-8pt] 
      &  $3z^2-r^2 $ &  $x^2-y^2 $  &  $3z^2-r^2 $ &  $x^2-y^2 $ &  $3z^2-r^2 $ &  $x^2-y^2 $ &  $3z^2-r^2 $ &  $x^2-y^2 $ \\ 
\hline \\ [-8pt] 
$3z^2-r^2 $  & 3.584 &  3.889  & 0.733  & 0.820 & 2.608  & 2.670 & 0.470 &  0.492 \\
$x^2-y^2 $  &   3.889 &  4.194 & 0.820  & 0.911 &  2.670 & 2.755 & 0.492 & 0.520 \\
\hline \\ [-8pt]
occ.(GWA)      &  $3z^2-r^2 $ &  $x^2-y^2 $  \\ 
\hline \\ [-8pt] 
                     & 1.992  &  1.008   \\
\hline
\hline 
\end{tabular} 
\end{table*} 

\begin{table*}[ptb] 
\caption{
Transfer integral and effective interaction in one-band Hamiltonian for HgBa$_2$CuO$_4$ (in eV). We show the transfer integral in the cGW and also in the GWA for comparison.
The transfer integrals denoted as GWA are calculated from the Wannier orbitals constructed to fit the GW band structure. The transfer integrals denoted as cGW-SIC are calculated from the cGW procedure applied to the GWA band structure/Green's functions.
The effective interaction is the result of the cRPA applied to the GWA Green's functions.
$v$ and $J_{v}$ represent the bare Coulomb interaction/exchange interactions respectively. $U(0)$ and $J(0)$ represent the static values of the effective Coulomb interaction/exchange interactions (at $\omega=0$). 
The data are the same as and taken from Table II of Ref.\onlinecite{hirayama18}.
}
\ 
\label{paraHg1v0} 
\begin{tabular}{c|c|c|c|c|c|c|c} 
\hline \hline \\ [-8pt]
$t $(GWA)   &     $(0,0,0)$  &   $(1,0,0)$ &     $(1,1,0)$ &   $(2,0,0)$     \\ [+1pt]
\hline \\ [-8pt] 
  $x^2-y^2$    &   0.164 & -0.453 &0.074 &  -0.051  \\
\hline \hline \\ [-8pt]
$t $(cGW)   &      $(0,0,0)$  &   $(1,0,0)$ &     $(1,1,0)$ &      $(2,0,0)$    \\ [+1pt]
\hline \\ [-8pt] 
$x^2-y^2$  & 0.190 &-0.461 &  0.119 & -0.072 \\
\hline \hline \\ [-8pt]  
               &      $v$  &     $U(0)$    \\ [+1pt]
\hline \\ [-8pt]
$x^2-y^2$   &  17.421  & 4.374  \\
\hline
\hline 
\end{tabular} 
\end{table*}

\begin{table*}[ptb] 
\caption{
Transfer integrals and effective interactions for one-band Hamiltonian of HgBa$_2$CuO$_4$ (in eV).
We show the transfer integrals obtained from the Wannier function constructed from the fitting to the GW+LRFB band structure, which are denoted as GW+LRFB. Transfer integrals obtained by the cGW procedure by using the  GW+LRFB bands and Green's functions are denoted by the cGW+LRFB. 
Effective interaction is obtained by using cRPA applied to the GW+LRFB Green's function.
$v$ represents the bare Coulomb interaction.
$U(0)$represent the static values of the effective Coulomb interaction (at $\omega=0$).
The index 'n' and 'nn' represent the nearest unit cell [1,0,0] and the next-nearest unit cell [1,1,0] respectively. 
The parameters for further neighbor transfer integrals and interactions by the cGW-SIC+LRFB are given in the Supplemental Material~\cite{SM}.
}
\ 
\label{paraHg1} 
\begin{tabular}{c|c|c|c|c|c|c|c} 
\hline \hline \\ [-8pt]
$t $(GW+LRFB)   &     $(0,0,0)$  &   $(1,0,0)$ &     $(1,1,0)$ &   $(2,0,0)$     \\ [+1pt]
\hline \\ [-8pt] 
  $x^2-y^2$    &   -0.111 & -0.512 & 0.082 &  -0.066  \\
\hline \hline \\ [-8pt]
$t $(cGW+LRFB)   &      $(0,0,0)$  &   $(1,0,0)$ &     $(1,1,0)$ &      $(2,0,0)$    \\ [+1pt]
\hline \\ [-8pt] 
$x^2-y^2$  & 0.229 & -0.509 &   0.127 & -0.077 \\
\hline \hline \\ [-8pt]  
               &      $v$  &     $U(0)$ &     $v_{\text{n}}$ &    $V_{\text{n}}(0)$ &   $v_{\text{nn}}$  &     $V_{\text{nn}}(0)$    \\ [+1pt]
\hline \\ [-8pt]
$x^2-y^2$   &  16.197  & 3.846 &  4.194  & 0.834 & 2.755  & 0.460 \\
\hline
\hline 
\end{tabular} 
\end{table*}

\begin{widetext}
\begin{table*}[h!] 

\caption{
Summary of effective Hamiltonian parameters for HgBa$_2$CuO$_4$ in the cGW-SIC+LRFB (in eV).
We show the transfer integral calculated from the GW as well as that calculated from the GW-SIC+LRFB procedure.
$t$ and $t'$ for one- and two-band Hamiltonians are for nearest and next nearest-neighbor transfers between Cu $3d$ orbitals, respectively.
Onsite and nearest-neighbor interactions $U$ and $V$, respectively for Cu $3d$ orbitals are given as well. 
The orbital level is given by $\epsilon_{X}$ with $X=x^2-y^2$ or $3z^2-r^2$. Left panel:1-band Hamiltonians. Middle two panels: two-band Hamiltonians. 
Right panel: three-band Hamiltonians 
$t_{dp}$ ($t_{pp}$) is for largest nearest-neighbor transfer between Cu $3d_{x^2-y^2}$ and O $2p_{\sigma}$ (two O $2p_{\sigma}$) orbitals.
Onsite ($U$) and nearest-neighbor ($V$) interactions for Cu $3d_{x^2-y^2}$ and O $2p_{\sigma}$ are given as well.
The level difference between $3d_{x^2-y^2}$ and $2p_{\sigma}$ is given by $\Delta\mu_{dp}$. 
}
\begin{tabular}{cccc}
\begin{minipage}{0.18\hsize}
\begin{center}
\
\begin{tabular}{c|c} 
\hline \hline \\ [-8pt]
  from GW & 1-band \\
\hline  \\ [-8pt]
$t$  &     -0.461    \\ [+1pt]
\hline \\ [-8pt] 
 $t'$    & 0.119    \\
\hline  \\ [-8pt]
$|t'/t|$  &  0.26     \\ [+1pt]
\hline \\ [-8pt] 
$U$  & 4.37 \\   [+1pt]
\hline \\ [-8pt] 
$V$  & 1.09 \\
\hline \\ [-8pt]  
 $|U/t|$   &  9.48    \\ 
\hline
\hline 
\\ [-8pt]
from GW+LRFB & 1-band \\
\hline  \\ [-8pt]
$t$  &     -0.509    \\ [+1pt]
\hline \\ [-8pt] 
 $t'$    & 0.127   \\
\hline  \\ [-8pt]
$|t'/t|$  &  0.25     \\ [+1pt]
\hline \\ [-8pt] 
$U$  & 3.85 \\   [+1pt]
\hline \\ [-8pt] 
$V$  & 0.83 \\
\hline \\ [-8pt]  
 $|U/t|$   &  7.56  \\
\hline
\hline 
\end{tabular} 
\end{center} 
\end{minipage} 
\begin{minipage}{0.29\hsize}
\begin{center}
\
\begin{tabular}{c|c|c} 
\hline \hline \\ [-8pt]
from GW & 2-band   \\ [+1pt]
\hline \\ [-8pt] 
  $t$    &  $3z^2-r^2 $ &  $x^2-y^2 $   \\ 
\hline \\ [-8pt] 
$3z^2-r^2 $  &  0.013 & 0.033  \\
$x^2-y^2 $  &  0.033  & -0.426  \\
\hline \\ [-8pt] 
  $t'$    &  $3z^2-r^2 $ &  $x^2-y^2 $  \\ 
\hline \\ [-8pt] 
$3z^2-r^2 $  &  -0.003 & 0.000  \\
$x^2-y^2 $  &  0.000  & 0.102  \\
\hline \\ [-8pt] 
  $|t_{x^2-y^2}'/t_{x^2-y^2}|$  & 0.24  \\ 
\hline \\ [-8pt] 
  $\epsilon_{x^2-y^2}-\epsilon_{3z^2-r^2}$ & 4.01  \\ 
\hline \\ [-8pt] 
  $U$    &  $3z^2-r^2 $ &  $x^2-y^2 $   \\ 
\hline \\ [-8pt] 
$3z^2-r^2 $  &  6.92 & 4.00 \\
$x^2-y^2 $  &  4.00  & 4.51  \\
\hline \\ [-8pt] 
  $V$    &  $3z^2-r^2 $ &  $x^2-y^2 $ \\ 
\hline \\ [-8pt] 
$3z^2-r^2 $  &  0.76 & 0.83  \\
$x^2-y^2 $  &  0.83  & 0.90  \\
\hline \\ [-8pt] 
  $|U/t_{x^2-y^2}|$    &  $3z^2-r^2 $ &  $x^2-y^2 $  \\ 
\hline \\ [-8pt] 
$3z^2-r^2 $  &  16.2 & 9.4  \\
$x^2-y^2 $  &  9.4  & 10.6 \\
\hline \hline 
\end{tabular} 
\end{center} 
\end{minipage} 
\begin{minipage}{0.29\hsize}
\begin{center}
\
\begin{tabular}{c|c|c} 
\hline \hline \\ [-8pt]
 from GW{+LRFB} & 2-band  \\ [+1pt]
\hline \\ [-8pt] 
  $t$    &  $3z^2-r^2 $ &  $x^2-y^2 $   \\ 
\hline \\ [-8pt] 
$3z^2-r^2 $  &  0.002 & 0.029  \\
$x^2-y^2 $  &  0.029  & -0.455  \\
\hline \\ [-8pt] 
  $t'$    &  $3z^2-r^2 $ &  $x^2-y^2 $  \\ 
\hline \\ [-8pt] 
$3z^2-r^2 $  &  -0.003 & 0.000  \\
$x^2-y^2 $  &  0.000  & 0.096  \\
\hline \\ [-8pt] 
  $|t_{x^2-y^2}'/t_{x^2-y^2}|$  &  0.21  \\ 
\hline \\ [-8pt] 
  $\epsilon_{x^2-y^2}-\epsilon_{3z^2-r^2}$ & 3.71  \\ 
\hline \\ [-8pt] 
  $U$    &  $3z^2-r^2 $ &  $x^2-y^2 $   \\ 
\hline \\ [-8pt] 
$3z^2-r^2 $  &  5.96 & 3.50 \\
$x^2-y^2 $  &  3.50  & 4.03  \\
\hline \\ [-8pt] 
  $V$    &  $3z^2-r^2 $ &  $x^2-y^2 $ \\ 
\hline \\ [-8pt] 
$3z^2-r^2 $  &  0.73 & 0.82  \\
$x^2-y^2 $  &  0.82  & 0.91 \\
\hline \\ [-8pt] 
  $|U/t_{x^2-y^2}|$    &  $3z^2-r^2 $ &  $x^2-y^2 $  \\ 
\hline \\ [-8pt] 
$3z^2-r^2 $  &  13.1 & 7.7  \\
$x^2-y^2 $  &  7.7  & 8.9 \\
\hline \hline 
\end{tabular} 
\end{center} 
\end{minipage} 
\begin{minipage}{0.22\hsize}
\begin{center}
\
\begin{tabular}{c|c} 
\hline \hline \\ [-8pt]
  from GW & 3-band \\
\hline  \\ [-8pt]
$t_{dp}$  &     1.257    \\ [+1pt]
\hline \\ [-8pt] 
 $t_{pp}$    & 0.751    \\
\hline  \\ [-8pt]
$\Delta_{dp}$  &  2.416     \\ [+1pt]
\hline \\ [-8pt] 
$U_{dd}$  & 8.84 \\   [+1pt]
\hline \\ [-8pt] 
$V_{dd}$  & 0.80 \\
\hline \\ [-8pt] 
$V_{dp}$  & 1.99 \\   [+1pt]
\hline \\ [-8pt] 
$U_{pp}$  & 5.31 \\
\hline \\ [-8pt]  
$V_{pp}$  & 1.21 \\
\hline \\ [-8pt]  
 $|U_{dd}/t_{dp}|$   &  7.03    \\ 
\hline
\hline 
\\ [-8pt]
from GW{+LRFB} & 3-band \\
\hline  \\ [-8pt]
$t_{dp}$  &     1.261    \\ [+1pt]
\hline \\ [-8pt] 
 $t_{pp}$    & 0.753    \\
\hline  \\ [-8pt]
$\Delta_{dp}$  &  2.174     \\ [+1pt]
\hline \\ [-8pt] 
$U_{dd}$  & 8.99 \\   [+1pt]
\hline \\ [-8pt] 
$V_{dd}$  & 0.84 \\
\hline \\ [-8pt] 
$V_{dp}$  & 2.05 \\   [+1pt]
\hline \\ [-8pt] 
$U_{pp}$  & 5.40 \\
\hline \\ [-8pt]  
$V_{pp}$  & 1.25 \\
\hline \\ [-8pt]  
$|U_{dd}/t_{dp}|$   &  7.13    \\ 
\hline
\hline 
\end{tabular} 
\end{center} 
\end{minipage} 
\end{tabular} 
\label{parameter_summary} 
\end{table*} 
\end{widetext}

%
\begin{table*}[h] 
\caption{
Transfer integrals and effective interactions for three-band Hamiltonian of La$_2$CuO$_4$ in the cGW-SIC (in eV) as well as in the GWA.
The notations are the same as Table~\ref{paraHg3_cGW-SIC}. The GWA and cGW-SIC data are taken from Table VII in Ref.\onlinecite{hirayama18}.
The parameters for further neighbor interactions by the cGW-SIC+$\Delta\mu$ are given in the Supplemental Material~\cite{SM}.
} 
\label{paraLa3_cGW-SIC} 
\begin{tabular}{c|ccc|ccc|ccc|ccc|ccc} 
\hline \hline \\ [-8pt]
$t $(GWA)   &       &  $(0,0,0)$  &       &     & $(1,0,0)$ &    &       & $(1,1,0)$ &      &     &  $(2,0,0)$ &     \\ [+1pt]
\hline \\ [-8pt]
      &  $x^2-y^2$ &  $p_1$ &  $p_2$ & $x^2-y^2$ &  $p_1$ &  $p_2$ &  $x^2-y^2$ &  $p_1$ &  $p_2$ & $x^2-y^2$ &  $p_1$ &  $p_2$ \\ 
\hline \\ [-8pt] 
$x^2-y^2$  & -1.743 & -1.399 &  1.399  & -0.010 & -0.012 & -0.042 &  0.013 & -0.006 &  0.006 &  -0.004 & -0.000 & -0.001  \\ 
$p_1$          & -1.399 & -4.657 & -0.659  &   1.399 & 0.120  & 0.659  & -0.042 & 0.041 & -0.000 & 0.012 & -0.002 & -0.000  \\
$p_2$          &  1.399 & -0.659 & -4.657  &  -0.042 & 0.000 & -0.011 &  0.042 &  -0.000 &  0.041   & -0.002 &   0.000 & -0.002    \\
\hline \hline \\ [-8pt]
$t $(cGW-SIC)   &       &  $(0,0,0)$  &       &     & $(1,0,0)$ &    &       & $(1,1,0)$ &      &     &  $(2,0,0)$ &     \\ [+1pt]
\hline \\ [-8pt]
      &  $x^2-y^2$ &  $p_1$ &  $p_2$ & $x^2-y^2$ &  $p_1$ &  $p_2$ &  $x^2-y^2$ &  $p_1$ &  $p_2$ & $x^2-y^2$ &  $p_1$ &  $p_2$ \\ 
\hline \\ [-8pt] 
$x^2-y^2$  & -1.538 & -1.369 &  1.369  & 0.038 & -0.036 & -0.028 &  0.025 & -0.020 &  0.020 & -0.005 &  0.005 &  0.005 \\ 
$p_1$          & -1.369 & -5.237 & -0.753  &  1.369 &  0.189 &  0.754  &  -0.028 &  0.047 &  0.010 & 0.036 & -0.005 &  0.009 \\
$p_2$          &  1.369 & -0.753 & -5.237  &  -0.029 & -0.010 &  0.021 &  0.028 & 0.009 &  0.047  & 0.005 & -0.002 &  0.002   \\
\hline \hline \\ [-8pt]
$t $(cGW-SIC$+\Delta\mu$)   &       &  $(0,0,0)$  &       &     & $(1,0,0)$ &    &       & $(1,1,0)$ &      &     &  $(2,0,0)$ &     \\ [+1pt]
\hline \\ [-8pt]
      &  $x^2-y^2$ &  $p_1$ &  $p_2$ & $x^2-y^2$ &  $p_1$ &  $p_2$ &  $x^2-y^2$ &  $p_1$ &  $p_2$ & $x^2-y^2$ &  $p_1$ &  $p_2$ \\ 
\hline \\ [-8pt] 
$x^2-y^2$  & -1.538 & -1.369 &  1.369  & 0.038 & -0.036 & -0.028 &  0.025 & -0.020 &  0.020 & -0.005 &  0.005 &  0.005 \\ 
$p_1$          & -1.369 & -2.737 & -0.753  &  1.369 &  0.189 &  0.754  &  -0.028 &  0.047 &  0.010 & 0.036 & -0.005 &  0.009 \\
$p_2$          &  1.369 & -0.753 & -2.737  &  -0.029 & -0.010 &  0.021 &  0.028 & 0.009 &  0.047  & 0.005 & -0.002 &  0.002   \\
\hline \hline \\ [-8pt]  
   &       &  $v$  &       &     & $U(0)$ &    &       & $J_{v}$ &      &     &  $J(0)$ &     \\ [+1pt]
\hline \\ [-8pt]
      &  $x^2-y^2$ &  $p_1$ &  $p_2$ & $x^2-y^2$ &  $p_1$ &  $p_2$ &  $x^2-y^2$ &  $p_1$ &  $p_2$ & $x^2-y^2$ &  $p_1$ &  $p_2$ \\ 
\hline \\ [-8pt] 
$x^2-y^2$ & 28.784  &  8.246 &   8.246   & 9.612 &  2.680 &   2.680  &            &  0.065 &   0.065  &           &   0.049 &  0.049  \\ 
$p_1$         &   8.246  & 17.777 &  5.501  &  2.680  & 6.128  &  1.861  &  0.065  &            &  0.036  & 0.049  &  -          &  0.019 \\
 $p_2$        &   8.246  &  5.501  & 17.777  & 2.680  &  1.861 &   6.128 &  0.065   & 0.036  &            &  0.049  & 0.019 &         \\
\hline \hline \\ [-8pt]  
       &       &  $v_{\text{n}}$ &    &     & $V_{\text{n}}(0)$ &    &       & $v_{\text{nn}}$  &      &     &  $V_{\text{nn}}(0)$ &     \\ [+1pt]
\hline \\ [-8pt] 
      &  $x^2-y^2$ &  $p_1$ &  $p_2$ & $x^2-y^2$ &  $p_1$ &  $p_2$ &  $x^2-y^2$ &  $p_1$ &  $p_2$ & $x^2-y^2$ &  $p_1$ &  $p_2$ \\ 
\hline \\ [-8pt] 
$x^2-y^2$ &   3.897 &  8.246  &  3.441  & 1.511  &  2.680  & 1.353   &  2.779  &  3.441  &  3.441 &  1.208  & 1.354  & 1.354  \\
$p_1$         &   2.656  & 4.002  &  2.502  &  1.199  &  1.503  &  1.156  &  2.241  & 2.770   & 2.502  &  1.104  & 1.217 &  1.157 \\
$p_2$        &   3.441  & 5.501  &  3.727   &  1.354  &  1.862  &  1.394  &  2.241  &  2.502 &  2.770  &  1.104  & 1.157  & 1.217 \\
\hline \\ [-8pt]
occ.(GWA)      &  $x^2-y^2$ &  $p_1$ &  $p_2$  \\ 
\hline \\ [-8pt] 
                     & 1.350 & 1.825  & 1.825   \\
\hline
\hline 
\end{tabular} 
\end{table*}

\begin{acknowledgments}
They are indebted to Takashi Miyake for his advice.
The authors acknowledge Terumasa Tadano, Yusuke Nomura and Kota Ido for useful discussions.
This work is financially supported by the MEXT HPCI Strategic Programs, and the Creation 
of New Functional Devices and High-Performance Materials to 
Support Next Generation Industries (CDMSI). This work was also supported 
by a Grant-in-Aid for Scientific Research (Nos. 16H06345 and 16K17746) from MEXT, Japan. 
TM was supported by Building of Consortia for the Development of Human Resources
in Science and Technology from the MEXT of Japan.
TO was supported by a Grant-in-Aid for Scientific Research No.18K13477.
The simulations were partially performed on the K computer 
provided by the RIKEN Advanced Institute for Computational Science 
under the HPCI System Research project (the project number hp170263 and hp180170). 
The calculations were also performed on computers at 
the Supercomputer Center, Institute for Solid State Physics, University of Tokyo.
\end{acknowledgments}
%
%
 



\clearpage

\renewcommand{\thetable}{S.\arabic{table}}
\setcounter{table}{0}
\renewcommand{\thesection}{S.\arabic{section}}
\renewcommand{\thefigure}{S.\arabic{figure}}


%
%
\begin{widetext}

\noindent
{\Large Supplementary material for Effective Hamiltonian for cuprate superconductors derived from multi-scale \textit{ab initio} scheme with level renormalization}

\noindent
\section*{S.1 Details of Hamiltonans}

In this supplementary material, we list up the whole parameters including relatively small one-body and two-body parameters. We show all the transfer integrals when they are above 10meV. Beyond the relative distance (3,3,0)  all the one-body parameters are below 10 meV.  We also show two-body parameters up to the distance (3,3,0). Within the distance (3,3,0), we list up interactions only when the value is above 50 meV.  Interactions for further neighbor unit-cell pairs very well follows $1/r$ dependence inferred from the list.
One-body parameters in the cGW-SIC+$\Delta\mu$ for the three-band hamiltonian of HgBa$_2$CuO$_4$ are listed in Table \ref{paraHg3muall} and the interaction parameters are given in Tables~
S.2, S.3, S.4, S.5, and S.6.
One-body parameters in the cGW-SIC+LRFB for the three-band hamiltonian of HgBa$_2$CuO$_4$ are listed in 
Table S.7 and the interaction parameters are given in Tables~
S.8, S.9, S.10, S.11, and S.12.
The two-band hamiltonian parameters in the cGW-SIC+LRF are listed in Tables S.13, S.14, S.15, and S.16.
In the same way, the one-band hamiltonian parameters are listed in Tables S.17 and S.18.
The hamiltonian parameters in the cGW-SIC+$\Delta\mu$ for the three-band hamiltonian of La$_2$CuO$_4$ are given in the same order in Tables S.19-S.23. Note that the unit cell of  La$_2$CuO$_4$ has two copper atoms in the $z$ direction.

\begin{table*}[h] 
\label{paraHg3muall} 
\begin{center}
\caption{
Transfer integrals in the cGW-SIC$+\Delta\mu$ for three-band hamiltonian of HgBa$_2$CuO$_4$ (in eV).
The inter-layer hopping is omitted because its energy scale is under 10 meV.
}
\ 
 
\end{center}
\end{table}

\clearpage

\begin{table}
\label{WrHg3all_interlayer} 
\begin{center}
\caption{(Continued from Table S4.)
Diagonal effective interactions in the cGW-SIC$+\Delta\mu$ for three-band hamiltonian of HgBa$_2$CuO$_4$ (in eV).
Notations are the same as Table S2.
}
\ 
%
 
\end{center}
\end{table*}

\begin{table*}[h] 
\label{paraHg3all} 
\begin{center}
\caption{
Transfer integrals in the cGW-SIC+LRFB for three-band hamiltonian of HgBa$_2$CuO$_4$ (in eV).
The inter-layer hopping is omitted because its energy scale is under 10 meV.
}
\ 
%
 
\end{tabular} 
\end{center}
\end{table}
\begin{table}
\label{WrHg2all_interlayer} 
\begin{center}
\caption{(Continued from Table S10.)
Diagonal effective interactions in the cGW-SIC+LRFB for three-band hamiltonian of HgBa$_2$CuO$_4$ (in eV).
Notations are the same as Table S8.
}
\ 
 
\end{center}
\end{table*}

\begin{table*}[h] 
\caption{
Transfer integrals and onsite potentials in the cGW-SIC+LRFB for the two-band hamiltonian of HgBa$_2$CuO$_4$ (in eV).
The inter-layer hopping except for that in $(0,0,1)$ is omitted because its energy scale is under 10 meV.
}
\ 
\label{paraHg2all} 
%
 
\end{center} 
\end{table}



\begin{table*}[h] 
\caption{
Transfer integrals in the cGW-SIC for three-band hamiltonian of La$_2$CuO$_4$ (in eV).
The inter-layer hopping is omitted because its energy scale is under 10 meV.
}
\ 
\label{paraLa3all} 
%
 
\end{table} 

\clearpage

\begin{table*}[h] 
\label{WrLa3all_offdiagonal3} 
\caption{(Continued from Table S22.)
Diagonal and off-diagonal effective interactions in the cGW-SIC for three-band hamiltonian of La$_2$CuO$_4$ (in eV).
Notations are the same as Table S20.
}
\ 
\begin{tabular}{c|c} 
\hline \hline \\ [-8pt]
  $(0,0,0)$  2  $(0,0,0)$  2  $(1.5,1.5,1)$  2  $(1.5,1.5,1)$  2   & 1.574   0.988   \\ 
  $(0,0,0)$  2  $(0,0,0)$  2  $(1.5,1.5,1)$  3  $(1.5,1.5,1)$  3   & 1.550   0.980   \\ 
  $(0,0,0)$  3  $(0,0,0)$  3  $(1.5,1.5,1)$  1  $(1.5,1.5,1)$  1   & 1.493   0.968   \\ 
  $(0,0,0)$  3  $(0,0,0)$  3  $(1.5,1.5,1)$  2  $(1.5,1.5,1)$  2   & 1.550   0.980   \\ 
  $(0,0,0)$  3  $(0,0,0)$  3  $(1.5,1.5,1)$  3  $(1.5,1.5,1)$  3   & 1.574   0.988   \\ 
  $(0,0,0)$  1  $(0,0,0)$  1  $(2.5,0.5,1)$  1  $(2.5,0.5,1)$  1   & 1.523   0.973   \\ 
  $(0,0,0)$  1  $(0,0,0)$  1  $(2.5,0.5,1)$  2  $(2.5,0.5,1)$  2   & 1.607   0.992   \\
\hline     
\hline 
\end{tabular} 
\end{table*}


\end{document}